\documentclass[12pt]{iopart}

\usepackage{enumitem}
\usepackage[caption=false]{subfig}
\usepackage{sidecap}
\usepackage{graphicx}
\usepackage[dvipsnames]{xcolor}
\usepackage{etoolbox}

\AtBeginEnvironment{figure}{\mathindent=0pt}
\AtBeginEnvironment{SCfigure}{\mathindent=0pt}

\usepackage{color}

\graphicspath{{images/}}

\begin{document}

\title[Controlling extended criticality]{Controlling extended criticality via modular connectivity}

\author{Nikita Gutjahr$^{1,2,*}$, Philipp Hövel$^{2,3,4,\dagger}$, and Aline Viol$^{2,4,5,\ddagger}$}

\address{$^1$ Institute of Theoretical Physics, Freie Universität Berlin, Arnimallee 14,
14195 Berlin, Germany}
\address{$^2$ Institute of Theoretical Physics, Technische Universität Berlin, Hardenbergstraße 36, 10623 Berlin, Germany}
\address{$^3$ School of Mathematical Sciences, University College Cork, Western Road, Cork, T12 XF62, Ireland}
\address{$^4$ Bernstein Center for Computational Neuroscience, Philippstraße 13, 10115 Berlin, Germany}
\address{$^5$ Cognitive Neuroscience, Scuola Internazionale Superiore di Studi Avanzati - SISSA, Trieste TS, 34136 Italy}
\ead{$^*$gutjahr.n@gmail.com}
\ead{$^\dagger$philipp.hoevel@ucc.ie}
\ead{$^\ddagger$aline.viol@sissa.it}


\begin{abstract}
Criticality has been conjectured as an integral part of neuronal network dynamics. Operating at a critical threshold requires precise parameter tuning and a corresponding mechanism remains an open question. Recent studies have suggested that topological features observed in brain networks give rise to a Griffiths phase, leading to power-laws in brain activity dynamics and the operational benefits of criticality in an extended parameter region. Motivated by growing evidence of neural correlates of different states of consciousness, we investigate how topological changes affect the expression of a Griffiths phase. We analyze the activity decay in modular networks using a Susceptible-Infected-Susceptible propagation model and find that we can control the extension of the Griffiths phase by altering intra- and intermodular connectivity. We find that by adjusting system parameters, we can counteract changes in critical behavior and maintain a stable critical region despite changes in network topology. Our results give insight into how structural network properties affect the emergence of a Griffiths phase and how its features are linked to established topological network metrics. We discuss how those findings can contribute to understand the observed changes in functional brain networks. Finally, we indicate how our results could be useful in the study of disease spreading.
\end{abstract}


\newpage

\section{\label{Intro}Introduction}
The criticality hypothesis states that biological neuronal networks are poised to operate at the critical threshold of a phase transition \cite{Beggs2008,  Chialvo2010, Tagliazucchi_2013, Mora2011}. It offers an explanation to characteristic scaling of power-law activity dynamics observed in such networks \cite{Beggs2003, Brochini2016, Haimovici2013}. This sheds light on the brain's information processing capabilities, as critical operation has been conjectured to optimize computational capability \cite{Bertschinger2004, Legenstein2007}, information transmission and storage \cite{Beggs2004, Haldeman2005, Lukovic2014}, and signal sensitivity and range \cite{Kinouchi2006, Shew2009}. While evidence for critical neuronal dynamics has been increasing \cite{Haimovici2013, Ma2019, Plenz2014}, an explanation  of how the brain could self-regulate at a precise critical point remains an open question  \cite{Beggs2012, Touboul2017}.

It has been shown that certain topological structures present in neuronal networks can cause the emergence of a critical region \cite{Munoz2010,Moretti2013}, substituting a single critical point, which would relax the necessity for fine-tuning parameters. Quenched disorder in networks can induce rare-region effects, resulting in critical behavior in an entire parameter region below the critical point, i.e., a \textit{Griffiths phase} \cite{Griffiths1969, Vojta2006}. Griffiths phases have been observed in synthetic hierarchical modular networks as well as in empirical and biologically inspired networks \cite{Moretti2013, Odor2015, Girardi_Schappo_2016}. Thereafter, it has been shown that sufficiently heterogeneous modular networks can support a Griffiths phase without hierarchy \cite{Cota2018}.

This study was driven by the following question: Given a self-regulating system poised at criticality, how would changes in its network topology affect its dynamics? Topological changes have been observed in functional brain networks of individuals in diverse states of consciousness, such as induced by psychedelics or anesthetics, in sleep or in coma \cite{Viol2017, Noirhomme2010, Schrouff2011, Andrade2011, Schroter2012, DeAraujo2012, Carhart-Harris2012, Palhano-Fontes2015}. The critical properties of a system are strongly determined by its topology \cite{David2006, Odor2004} and changes in topology can lead to an altered critical point. If the topology of a self-regulating system that only operates at criticality is changed, it would be forced to adapt. Maintaining critical operation could be achieved by either adjusting its parameters to the altered critical region, e.g. the rate of activity spread in brain networks, or by modifying its structure to revert the critical region to its previous parameter range.

In this paper, we investigate how topological properties influence dynamical processes in modular networks featuring a Griffiths phase. We study which network features are responsible for the emergence of a Griffiths phase and how one can manipulate its properties. We find a connection between the Griffiths phase width, i.e., the range of system parameter values that lead to power-law decay, and the network's topological properties. In short, we find that the Griffiths phase can persist in a changing topology and its width can be controlled via both intra- and intermodular connectivity. Alterations in the critical region that stem from a change in either connectivity can be counteracted by tuning the opposing structure. We argue that this could provide a mechanism of self-regulation in modular systems that operate at criticality and add to the functional benefits of modularity.

Our results give insight into how the structural properties of modular networks lead to the emergence of a Griffiths phase and connect it to established topological metrics. We highlight the importance of low global efficiency and propose that it is a central feature in this context. We suggest further inquiry into other artificial modular networks or real-world networks, such as empirical brain networks.

Finally, we discuss how an altered Griffiths phase could be connected to the topological changes observed in functional brain networks during altered states of consciousness. If consciousness relates to critical operation, could an increase in Griffiths phase width be connected to the reported changes in conscious quality under the influence of mind altering substances? Alongside, we briefly discuss how our results can help in understanding the persistence of a pandemic disease.

This paper is structured as follows: In the Methods section, we introduce the modular networks, the epidemiological model and our approach to analyze the Griffiths phase. In the Results section, we visualize the topological disorder in the modular networks. We show how a change in inter- and intramodular connectivity affects the Griffiths phase and topological network metrics. We conclude with a discussion of our results.

\begin{figure}
    \centering
	\subfloat[\label{fig:mmn}]{%
		\includegraphics[width=0.33\linewidth]{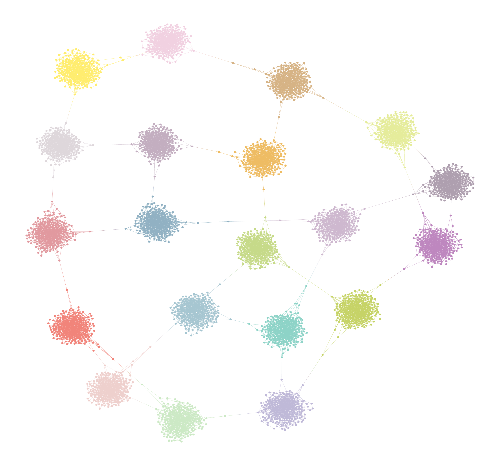}%
	}
	\subfloat[\label{fig:mmn_2}]{%
		\includegraphics[width=0.33\linewidth]{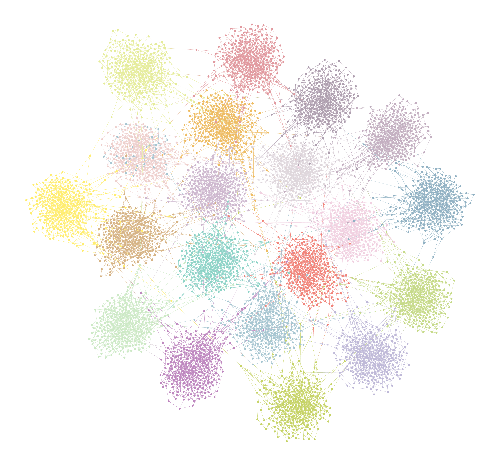}%
	}
	\subfloat[\label{fig:mmn_3}]{%
		\includegraphics[width=0.33\linewidth]{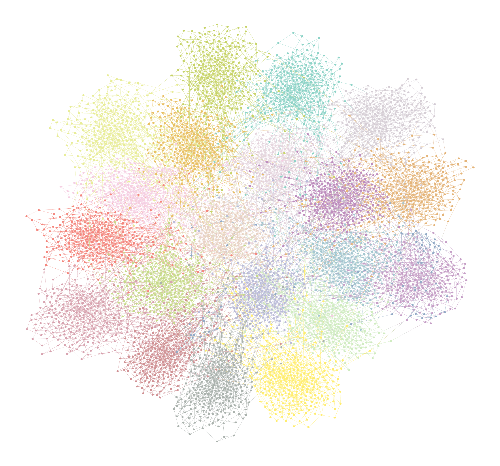}%
	}
	\caption{Schematic modular networks, showing two levels of distinct topological structure; ${M=20}$ modules of size ${N=10^3}$ are drawn from a power-law degree distribution with exponent ${\gamma=2.7}$ and connected via a regular intermodular network with (a) ${k_{\mathrm{inter}} = 3}$, (b) ${k_{\mathrm{inter}} = 11}$ and (c) ${k_{\mathrm{inter}} = 38}$ intermodular links per module. With increasing intermodular connectivity the modular networks converge to a non-modular power-law structure.}
	\label{fig:mmn_example}
\end{figure}

\section{Methods}

We explore how topological changes influence the Griffiths phase by simulating the Susceptible-Infectious-Susceptible (SIS) propagation model \cite{Hinrichsen2000, Pastor-Satorras2001} on synthetic modular networks. We choose the modular topology introduced in \cite{Cota2018} because, to our knowledge, it is the simplest modular structure reported in the literature that induces extensive Griffiths phase effects with different propagation models. The modular networks consist of a loosely connected ensemble of modules and offer a direct way to manipulate intra- and intermodular connectivity individually. An illustration of the networks can be seen in Figure~\ref{fig:mmn_example}.

\subsection{Constructing the modular networks}

We construct the networks by generating and randomly interconnecting $M$ modules of size $N$, each module drawn from the same power-law degree distribution $P_{\mathrm{intra}}(k) \sim k^{-\gamma}$. For simplicity, we generate the networks with modules of equal size and intermodular connections, leading to a random regular intermodular structure of degree $k_{\mathrm{inter}}$. This architecture is referred to as a \textit{monodisperse modular network} \cite{Cota2018}. Detailed instructions to generate the networks are given in \ref{appendix_networks}.

We consider networks with ${M=1000}$ modules consisting of ${N=1000}$ nodes each. The minimum degree in each module is ${k_{\mathrm{min}}=3}$, and the cut-off is set to ${k_{\mathrm{max}}}$, corresponding to the average maximal degree ${\langle k_{\mathrm{max}}(N) \rangle  \sim N^{\frac{1}{\gamma - 1}}}$ of a power-law network generated by the configuration model \cite{Catanzaro2005}. This cut-off leads to a distribution of critical points in individual module realizations, creating topological disorder within the modular networks. A discussion of how the cut-off choice impacts SIS dynamics in power-law networks is given in \cite{Cota2016}.

\subsection{Dynamical spreading process}
For the activity density decay analysis, we utilize the SIS spreading process. It was originally introduced as an epidemiological model for diseases that do not confer any immunity \cite{Anderson_May_1991}. A population is compartmentalized into susceptible and infectious members. Infectious members spread a disease to susceptible members with rate $\lambda$ and recover with rate $\mu$. After recovery an infectious member is again susceptible to reinfection. The SIS model features an absorbing state phase transition: A \textit{critical spreading rate} $\lambda_c$ separates a stationary from an absorbing phase. Above $\lambda_c$, the system converges to a stable density of infected/active members $\rho$. Below $\lambda_c$, the disease/activity eventually dies out. Due to its minimal assumptions, the SIS process is readily applicable in contexts that go beyond epidemiology, such as the spread of computer viruses or, as in our case, the activity in neuronal networks.

In a network model each node represents a member of the population and infected nodes spread activity to every susceptible neighbor node, which results in a high susceptibility to degree variations. The SIS process is a continuous-time Markov chain and its dynamics in a network can be simulated with the statistically exact Gillespie algorithm \cite{Bogu__2014}. In the present study, we use an optimized version of the algorithm that reduces the computational load of the simulation \cite{Cota2017}. Our implementation follows the description in \cite{Cota2018} and is detailed in \ref{appendix_SIS}.

\subsection{Susceptibility}
An important quantity in the analysis of complex, coupled systems is the \textit{susceptibility}. It diverges when a system undergoes a phase transition in dynamical spreading processes and can be utilized to calculate the critical threshold. We consider the susceptibility defined as follows \cite{Ferreira2012, Binder2010}:

\begin{equation*}
\chi = N \frac{\langle \rho^{2} \rangle - \langle \rho \rangle^2}{\langle \rho \rangle}.
\end{equation*}
The average activity density ${\langle \rho \rangle}$ was computed via the quasistationary (QS) method, where the system is kept in a QS state by returning it to a previous state whenever the activity dies out \cite{DeOliveira2005}. During the initial $m$ time steps of a simulation the state of the system is saved.
At each subsequent time step, a randomly chosen saved system state is overwritten by the current state with probability $p_{\mathrm{QS}}$. If the process reaches the absorbing state with ${N_{\mathrm{inf}} = 0}$, the system is returned to a randomly chosen saved state.

We used ${m = 70}$ saved states and an overwriting probability of ${p_{\mathrm{QS}} = 0.01}$, for which the system converges to a QS state. Then, the $n$-th moment of activity density is estimated by taking the respective temporal average of the steady state
\begin{equation}
    \langle \rho (t)^n \rangle = \frac{1}{T}\sum_{t=t^{'}}^{T} \rho (t)^n,
\end{equation}
where $T$ denotes the observation period, and $t^{'}$ is set large enough to discard the initial dynamical transient before the QS state is reached.

\subsection{Activity density decay analysis}
We study the Griffiths phase by performing an activity density decay analysis, as described in the following. Starting with a fully active network, we monitor the density of active nodes $\rho$ over time, averaged over multiple runs and network realizations. We use the spreading rate $\lambda$ as a control parameter and $\rho$ serves as the order parameter. By exploring a range of $\lambda$ values we observe an extended region showing power-law decay of $\rho(t)$ in the transition from subcritical absorbing states - characterized by exponential decay - to supercritical steady states. This extended region of slow decay is the signature of a present Griffiths phase \cite{Moretti2013, Cota2018}. 

The range $\Delta \mathrm{GP}$ of the power-law decay region is determined by the margin between the critical point $\lambda_{\mathrm{c}}$, defined as the highest value of $ \lambda$ that does not lead to a steady state, and the lowest spreading rate that shows power-law decay $\lambda_{\mathrm{low}}$, determined by the topological properties of the modules: 
\begin{equation}
 \Delta \mathrm{GP} = \lambda_{\mathrm{c}} - \lambda_{\mathrm{low}}.
\end{equation}
Increasing the intramodular connectivity of a modular network to the limit at which it becomes non-modular, its critical point converges to the lower boundary of the Griffiths phase ${\lambda_{\mathrm{c}}\approx\lambda_{\mathrm{low}}}$, annulling any Griffiths phase effects. We therefore identify $\lambda_{\mathrm{low}}$ as the critical point of this non-modular network that has the same structure as a single module, but the size of all modules combined.

\subsection{Determining $\lambda_{\mathrm{c}}$ and $\lambda_{\mathrm{low}}$}
Susceptibility diverges when a system undergoes phase transition in dynamical spreading processes and can be utilized to calculate the critical threshold \cite{Binder2010}. A Griffiths phase is accompanied by an extended region of high susceptibility, which makes this approach challenging. We therefore measure $\lambda_{\mathrm{c}}$ by increasing $\lambda$ in the density decay simulations until a value is reached that shows a steady state, and is therefore above the critical point. $\lambda_{\mathrm{c}}$ is then taken as the value midway between the first value above the critical point and the last value below it. The range between these rates is taken as the error. It should be noted that the error is not a standard deviation, since the likelihood of finding the true critical point within the error region does not have a Gaussian profile.

At the lower end of the Griffiths phase the power-law decay transitions into exponential decay continuously. To determine a distinct $\lambda_{\mathrm{low}}$ that separates the regions within and below the Griffiths phase, we additionally take into account how the decay behavior is influenced by changes in intermodular connectivity. By increasing the intermodular connectivity, we lower $\lambda_{\mathrm{c}}$ until we reach a non-modular structure that is equivalent to a single module of size ${M\cdot N}$ and has a critical point $\lambda_{\mathrm{c'}}$. The decay behavior for any $\lambda$ below $\lambda_{\mathrm{c'}}$ is not significantly affected by the change in intermodular connectivity (see supplementary material). However, any $\lambda$ above $\lambda_{\mathrm{c'}}$, that lies in the Griffiths phase at low $k_{\mathrm{inter}}$, will lead to a steady state when $k_{\mathrm{inter}}$ is increased. Therefore ${\lambda_{\mathrm{c'}}=\lambda_{\mathrm{low}}}$ is the natural lower limit of the Griffiths phase. 

In short, to determine $\lambda_{\mathrm{low}}$, we generate non-modular networks of size ${M \cdot N}$ and determine their critical point via susceptibility peaks. Note that the modular networks can be seen as diluted power-law networks, similar to the diluted Ising lattice, in which the Griffiths phase was originally proposed \cite{Griffiths1969}.

\subsection{Averaging over multiple networks}
The intermodular links are assigned at random in each network realization, which leads to a slightly varying $\lambda_{\mathrm{c}}$ and differing decay behavior, when $k_{\mathrm{inter}}$ is increased. Above ${k_{\mathrm{inter}} = 10}$ it is necessary to average over multiple networks to observe consistent power-law decay within the Griffiths phase. However, by increasing the number of modules to ${M=10^4}$ we can observe a consistent Griffiths phase in single network realizations up to ${k_{\mathrm{inter}} = 100}$. If we increase the intermodular connectivity beyond ${k_{\mathrm{inter}} = 100}$ the decay transitions into the decay of a non-modular power-law network even for very large modular networks. Further details can be found in Figure~\ref{fig:sup_decay} of the supplementary material.

\subsection{Topological metrics}
The structural properties of the modular networks change when generated in different configurations. To characterize these changes we utilize various topological metrics. Global efficiency \cite{Rubinov2010} is defined as

\begin{equation}
E = \frac{1}{N(N-1)} \sum_{i \neq j \in G} \frac{1}{d(i,j)},
\end{equation}
with total number of nodes $N$ and geodesic distance ${d(i,j)}$ from node $i$ to $j$ in graph $G$. It measures how well information can be exchanged within a network. It scales inverse to the characteristic path length and is high in integrated networks with low diameter and low when a network is segregated.

We also calculate the geodesic entropy of the networks. Geodesic entropy \cite{Viol2019} is a measure for how distributed the geodesic distances for a given node to all other nodes are. We calculate it by first determining the probabilities $p_i(r)$ that a given node ${i \in G}$ is of distance ${d(i,j) = r}$ to a randomly chosen node ${j \in \bar{G}}$ with ${\bar{G}=\{ j|j \in G \setminus \{i\} \}}$

\begin{equation}
    p_i(r) = \frac{1}{N-1}\sum_{j\in \bar{G}}\delta_{d(i,j),r},
\end{equation}
where $r$ lies in the interval ${1 \leq r \leq r_{\mathrm{max}}}$ with ${r_{\mathrm{max}} = \mathrm{max}_{j \in \bar{G}}(d(i,j))}$. The geodesic entropy is then given by

\begin{equation}
    s_i[p_i] = -\sum^{r_{\mathrm{max}}}_{r=1}p_i(r)~\mathrm{log}~p_i(r)
\end{equation}
and the characteristic geodesic entropy by taking the average over all nodes

\begin{equation}
    S_{\mathrm{geo}} = \sum^{N}_{i=1}s_i.
\end{equation}
The geodesic entropy allows us to quantify entropic changes due to structural properties, such as altering the intermodular connectivity in the modular networks. This leaves the degree distribution unchanged and can not be measured in the entropy of the degree distribution.

To measure entropic changes connected to the intramodular connectivity, we use the degree entropy

\begin{equation}
    S_{\mathrm{deg}}[P(k)] = -\sum^{k_{\mathrm{max}}}_{k=1}P(k)~\mathrm{log}~P(k).
\end{equation}
Another quantity we evaluate is the extended, local clustering coefficient \cite{Abdo2006}. It reveals neighbor relations that go beyond direct connectivity:
\begin{equation}
c^d_i = \frac{\left|\{\{u,v\};u,v \in N_i | d_{\bar{G}} (u,v) = d\}\right|}{{|N_i|\choose 2}},
\end{equation}
with the set of neighbors $N_i$ of node $i$. It measures the ratio between the number of pairs in $N_i$ whose distance is $d$ in ${G(V\setminus \{i\})}$ and the total number of pairs of neighbors. At ${d=1}$ it returns the standard clustering coefficient. We calculate the extended clustering coefficient with a function from the graph-tool library in python \cite{peixoto_graph-tool_2014}.

\section{\label{Results}Results}

\subsection{Module susceptibility}

\begin{figure}
    \centering
    \subfloat[\label{fig:crit_fluct_pl0}]{%
		\includegraphics[width=0.5\linewidth]{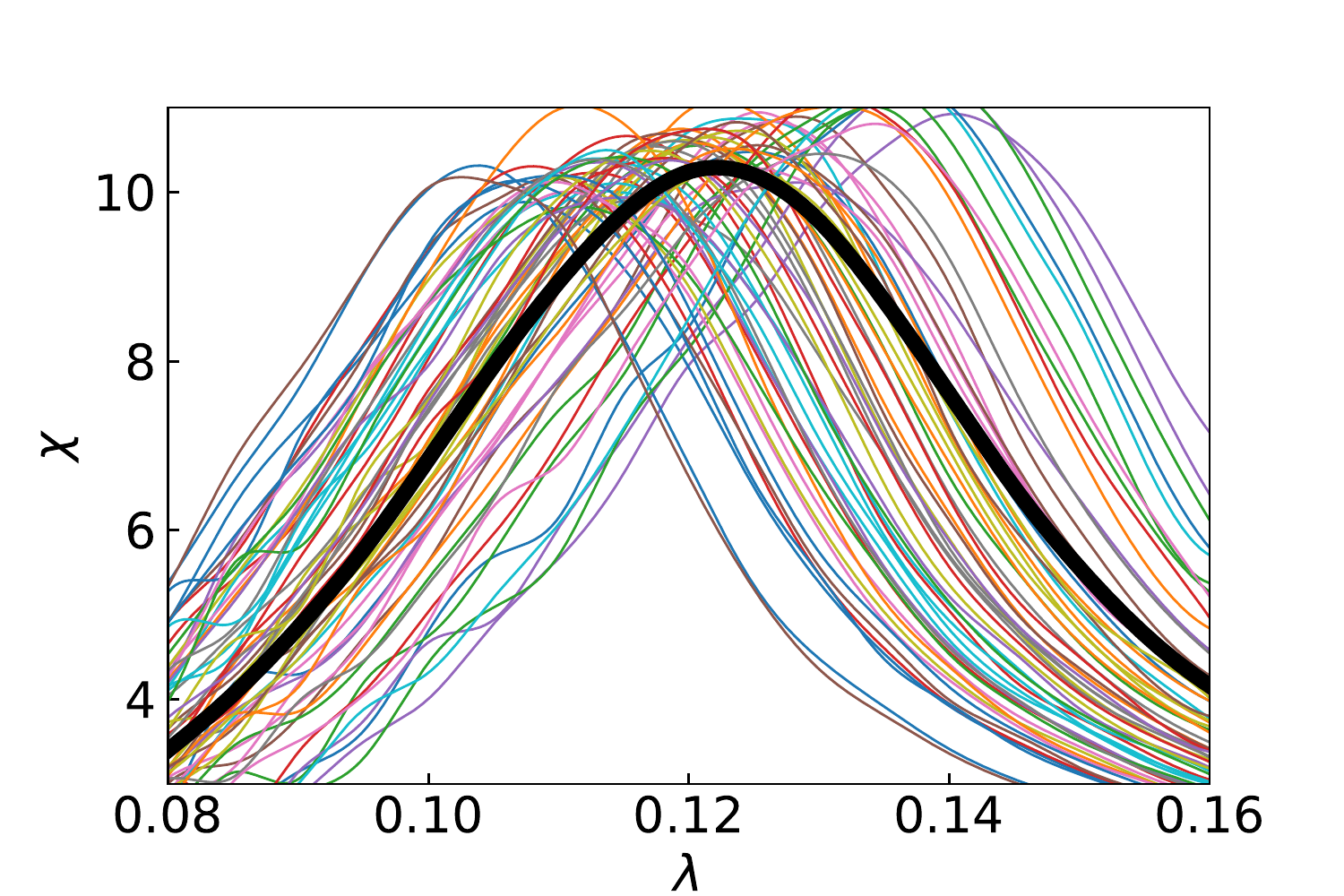}%
	}
	\subfloat[\label{fig:crit_fluct_pl2}]{%
		\includegraphics[width=0.5\linewidth]{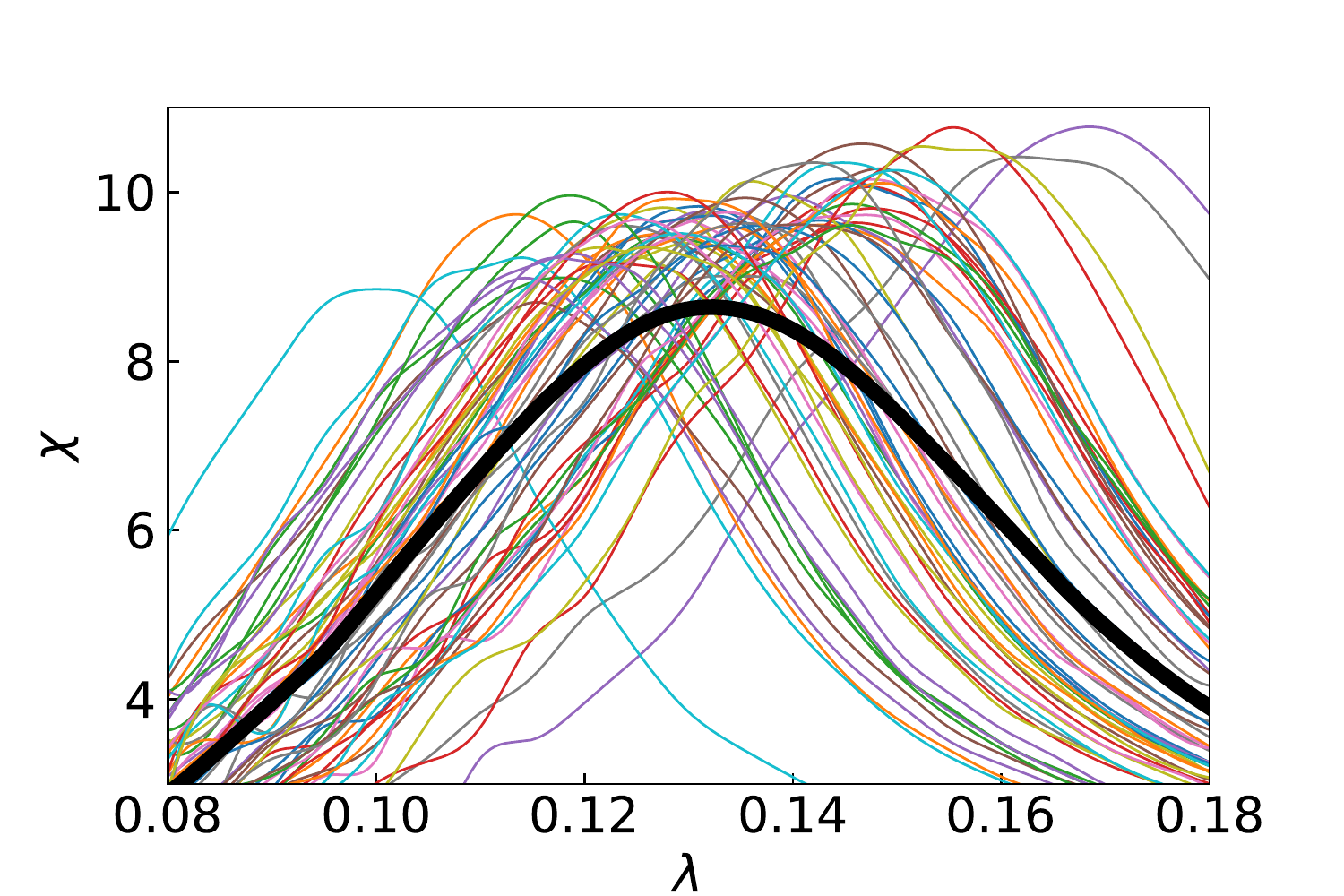}%
	}
	\caption{Dynamic susceptibility curves of 60 individual modules of size ${N=10^3}$ with ${\gamma = 2.7}$ (colored curves) and their average (black curve). The intramodular connectivity in (b) is reduced by subtracting 2 from the degree of each node during the generation of the module.}
	\label{fig:crit_fluct}
\end{figure}

Figure~\ref{fig:crit_fluct} shows the susceptibility of individual network modules. Individual module realizations have large variations in the number and connectivity of high degree outlier nodes \cite{Cota2016}. This leads to varying critical points with the SIS, which can be seen in shifting peaks of dynamical susceptibility. An extended region of high susceptibility is revealed by averaging the susceptibility of an ensemble of modules. This expresses how locally varying dynamics, caused by topological disorder, build the basis of the Griffiths phase.

\subsection{Changing the intermodular connectivity}

\begin{figure}
    \centering
	\subfloat[\label{fig:gp_cross1}]{%
	\includegraphics[width=0.49\linewidth]{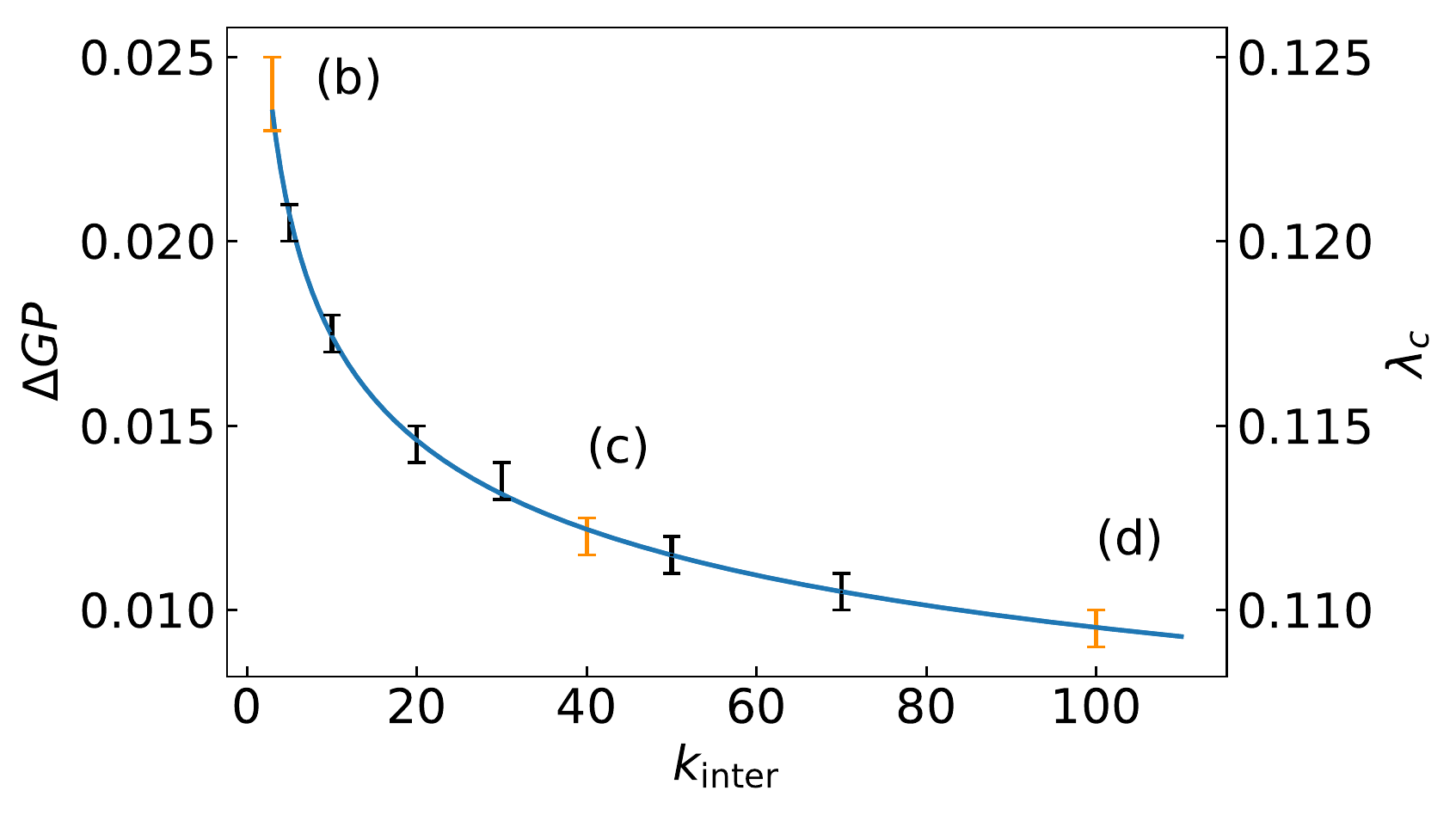}%
    }\hfill\\
	\subfloat[$k_{\mathrm{inter}}=3$\label{fig:mod_c3_decay}]{%
		\includegraphics[width=0.33\linewidth]{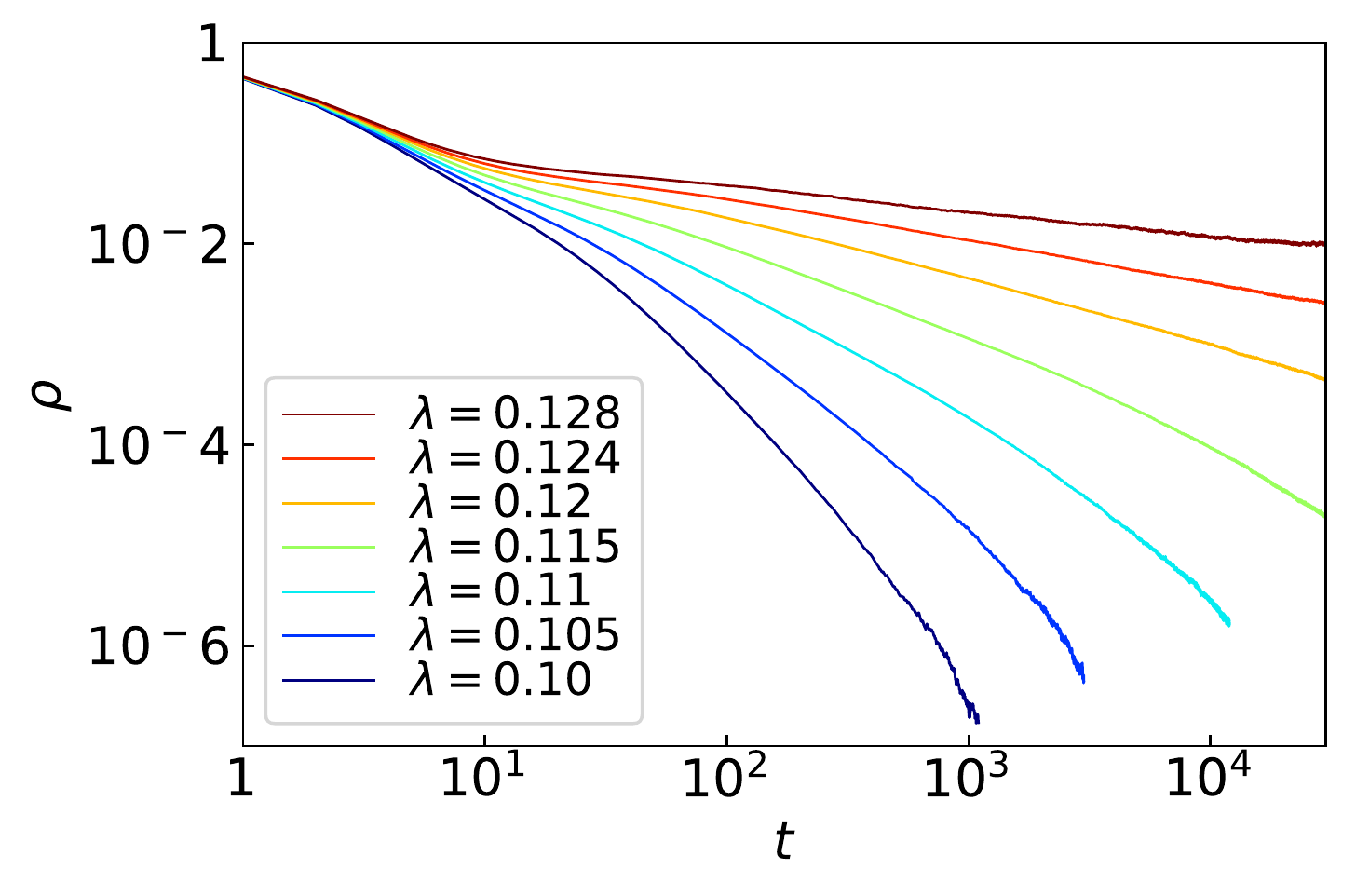}%
	}\hfill
	\subfloat[$k_{\mathrm{inter}}=40$\label{fig:mod_c40_decay}]{%
		\includegraphics[width=0.33\linewidth]{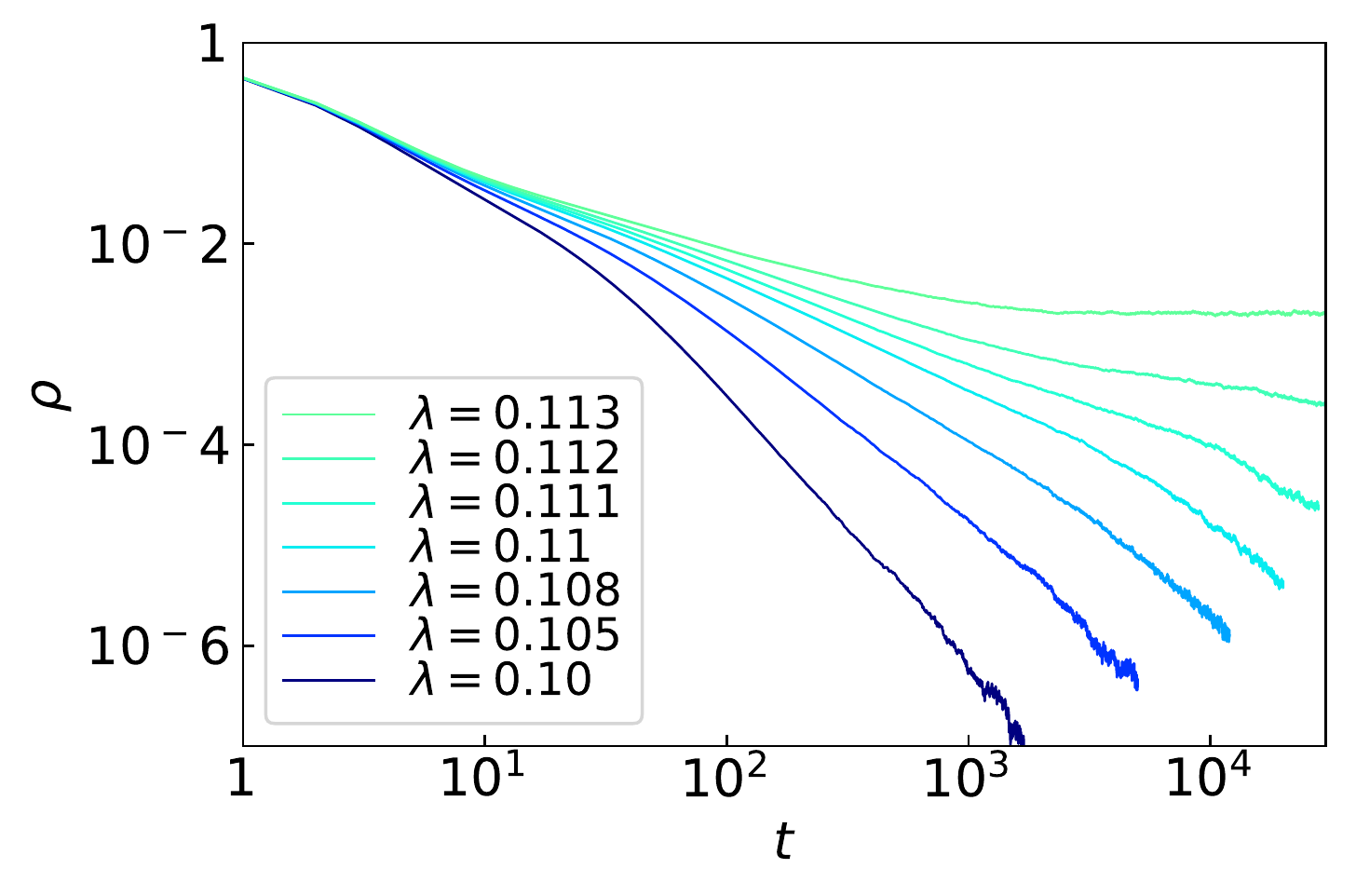}%
	}\hfill
	\subfloat[$k_{\mathrm{inter}}=100$\label{fig:mod_c100_decay}]{%
		\includegraphics[width=0.33\linewidth]{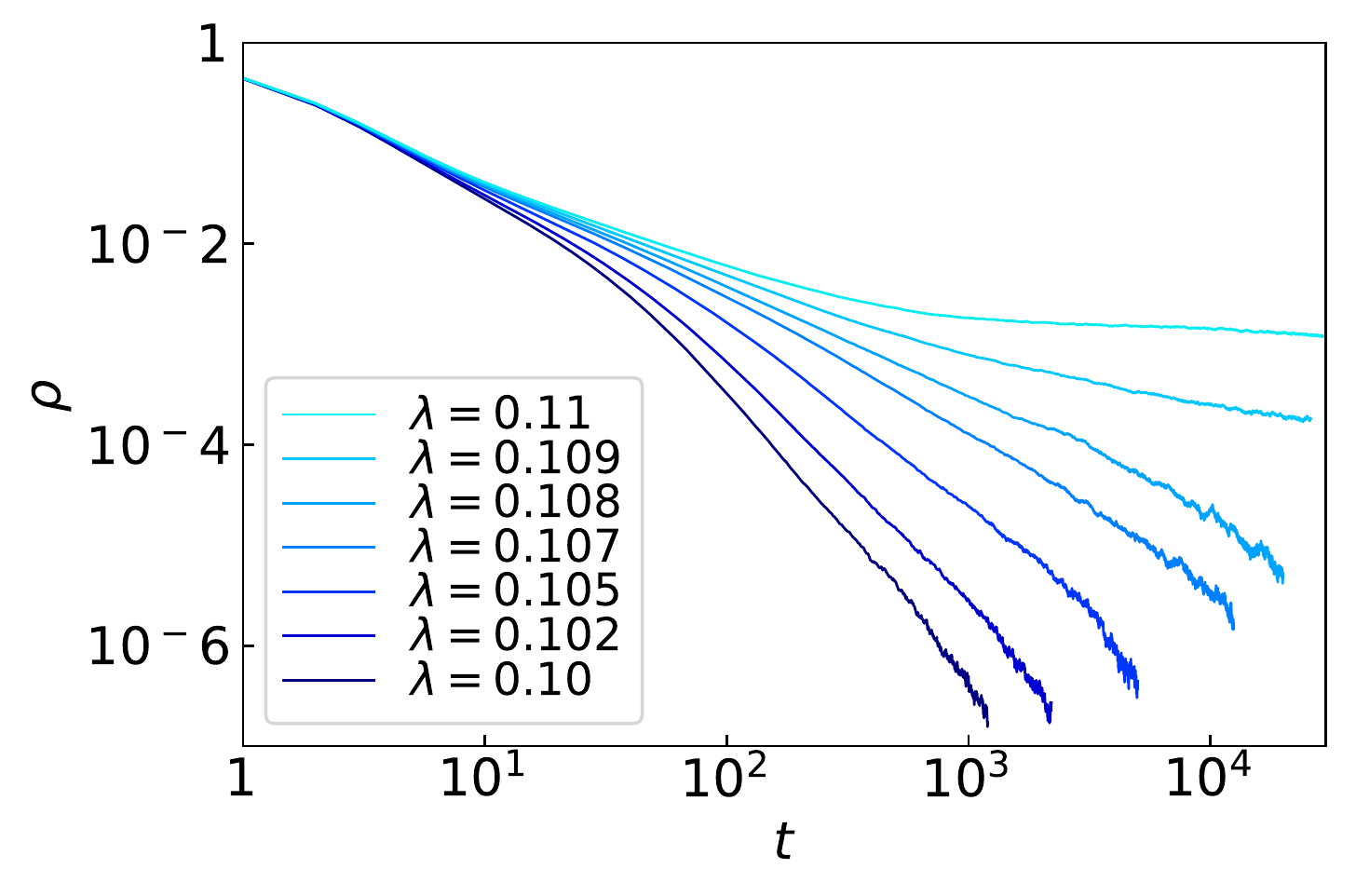}%
	}\hfill
	\caption{(a) The increase in $k_{\mathrm{inter}}$ leads to a decline in $\lambda_{\mathrm{c}}$ and $\Delta \mathrm{GP}$, as the power-law decay below $\lambda_{\mathrm{c}}$ remains unchanged, with ${\lambda_{\mathrm{low}}=0.10}$. (b-d) Activity density decay simulations in modular power-law networks with exponent ${\gamma = 2.7}$ with ${M=N=10^3}$ modules and nodes with equal intra- and increasing intermodular connectivity, averaged over ${7-350}$ runs in 10 networks each. The lowest decay lines correspond to $\lambda_{\mathrm{low}}$. $\lambda_{\mathrm{c}}$ is located between the two highest decay lines.}
	\label{fig:mod_decay_inter}
\end{figure}

Figure~\ref{fig:gp_cross1} shows how the Griffiths phase width depends on intermodular connectivity. We randomly distribute a fixed number $k_{\mathrm{inter}}$ of new intermodular connections per module, while keeping the degree distribution and intramodular connectivity unchanged. The number of added intermodular links is small enough to not significantly change the networks modularity.

The increase in $k_{\mathrm{inter}}$ leads to a reduction of the Griffiths phase width $\Delta \mathrm{GP}$.
Figure~\ref{fig:mod_decay_inter}b-d display detailed activity density decay simulations for modular networks with different numbers of intermodular connections (highlighted in Figure~\ref{fig:gp_cross1}). The reduction of $\Delta \mathrm{GP}$ stems from the reduction of $\lambda_{\mathrm{c}}$, since  $\lambda_{\mathrm{low}}$ remains constant.

\subsection{Changing the intramodular connectivity}

Introducing or removing intermodular links has a consistent influence on dynamic behavior, because of the regular intermodular structure. The influence of intramodular links depends on where the links are attached to: High-degree nodes have a stronger individual influence on the SIS dynamics than low-degree nodes. We therefore lower intramodular connectivity via two approaches: The first is to reduce each node degree ${k > 3}$ by a constant value, maintaining the minimal degree of ${k_{\mathrm{min}} = 3}$, which creates a shift in the degree distribution. We named this approach the \textit{offset method}. This method affects every node in the network and since most nodes are of low degree, it changes the intramodular connectivity via the low-degree nodes.

\begin{figure}
    \centering
    \subfloat[\label{fig:gp_decay}]{%
	    \includegraphics[width=0.49\linewidth]{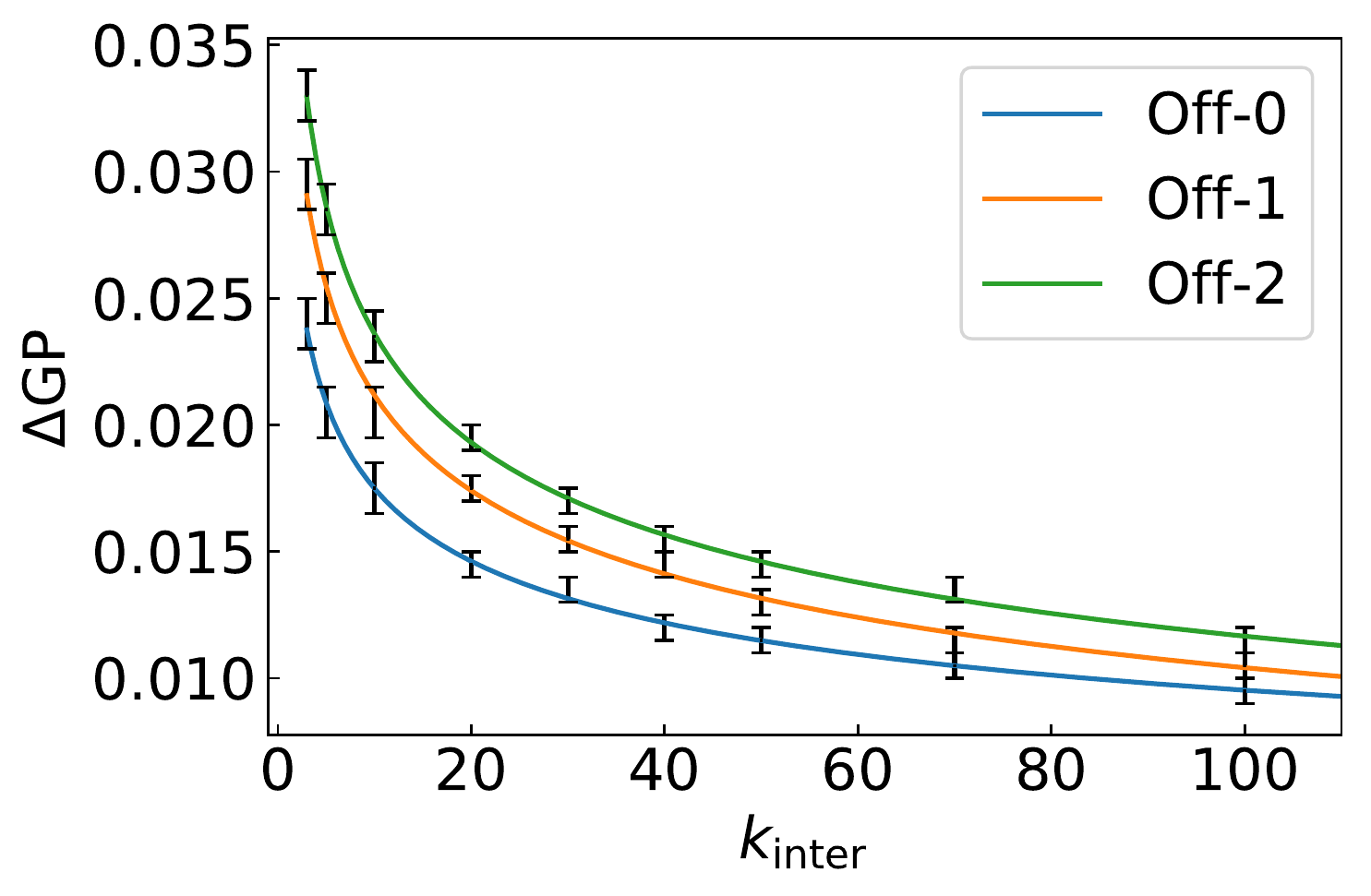}%
    }\hfill
	\subfloat[\label{fig:lc_decay}]{%
		\includegraphics[width=0.49\linewidth]{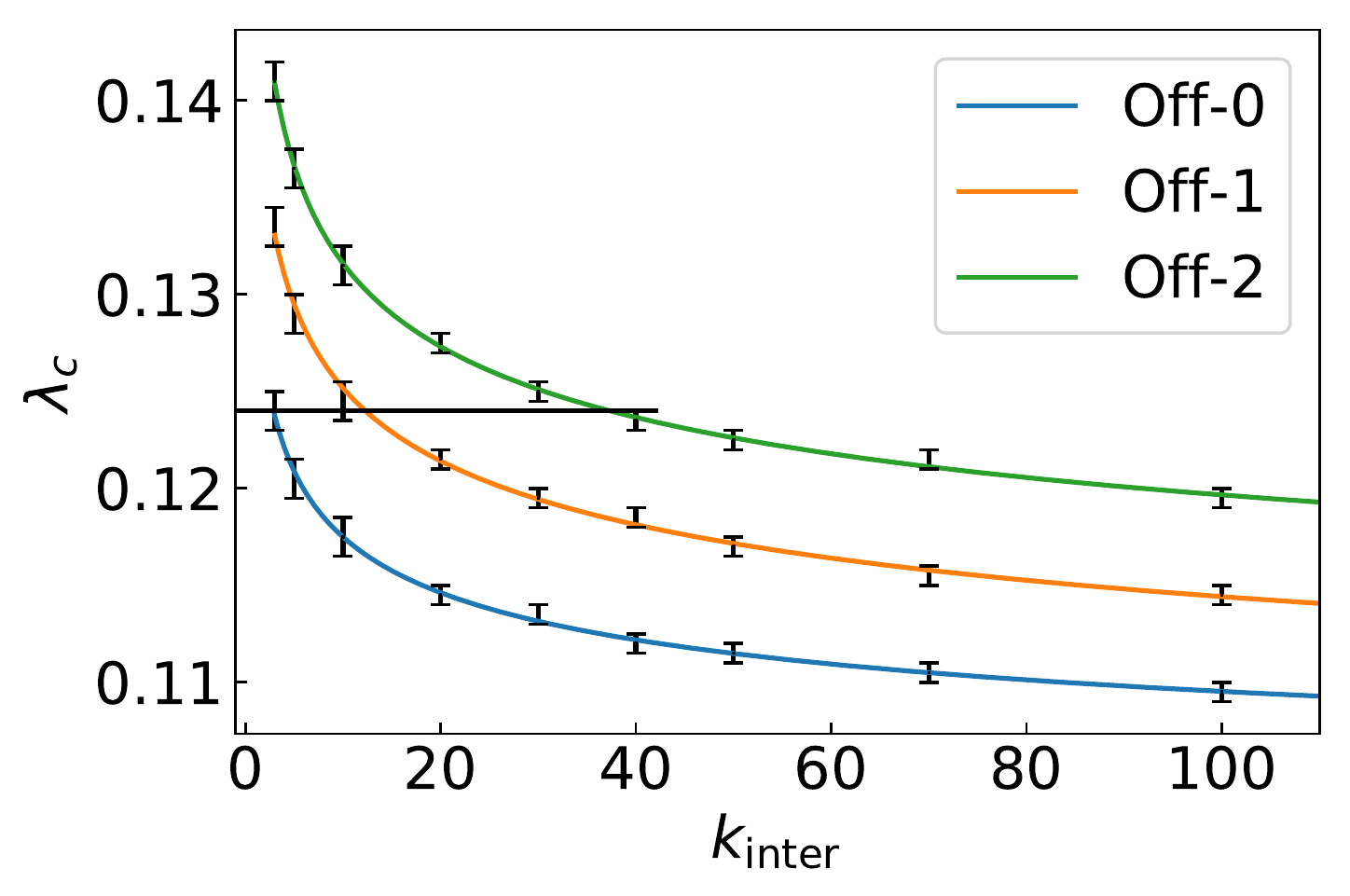}%
	}
	\caption{(a) $\Delta$GP and (b) $\lambda_{\mathrm{c}}$ reduction in modular power-law networks with exponent ${\gamma = 2.7}$ with ${M=N=10^3}$ modules and nodes with increasing intermodular connectivity at different levels of intramodular connectivity reduction. Introducing a degree offset shifts the decline of  $\Delta$GP and $\lambda_{\mathrm{c}}$ into a different parameter region. Density decay simulations for equal $\lambda_{\mathrm{c}}$ along the black line are shown in Figure~\ref{fig:counter_tuning}.}
	\label{fig:mod_decay_intra}
\end{figure}

The reduction of intramodular connectivity via the offset method increases $\lambda_{\mathrm{c}}$, such that the whole curve $\lambda_{\mathrm{c}}$ versus $k_{\mathrm{inter}}$ moves upwards (cf. Figure~\ref{fig:mod_decay_intra}). $\lambda_{\mathrm{low}}$ is also increased, but less than $\lambda_{\mathrm{c}}$ which leads to an increase in $\Delta$GP with lowered intramodular connectivity. One can observe that at low intermodular connectivity, $\lambda_{\mathrm{c}}$ and $\Delta \mathrm{GP}$ are more susceptible to changes in inter- than intramodular links. This behavior can be better understood by considering how topological metrics are affected by the changes in connectivity, in particular global efficiency.

\begin{figure}
    \centering
    \subfloat[\label{fig:crit_over_off}]{%
		\includegraphics[width=0.5\linewidth]{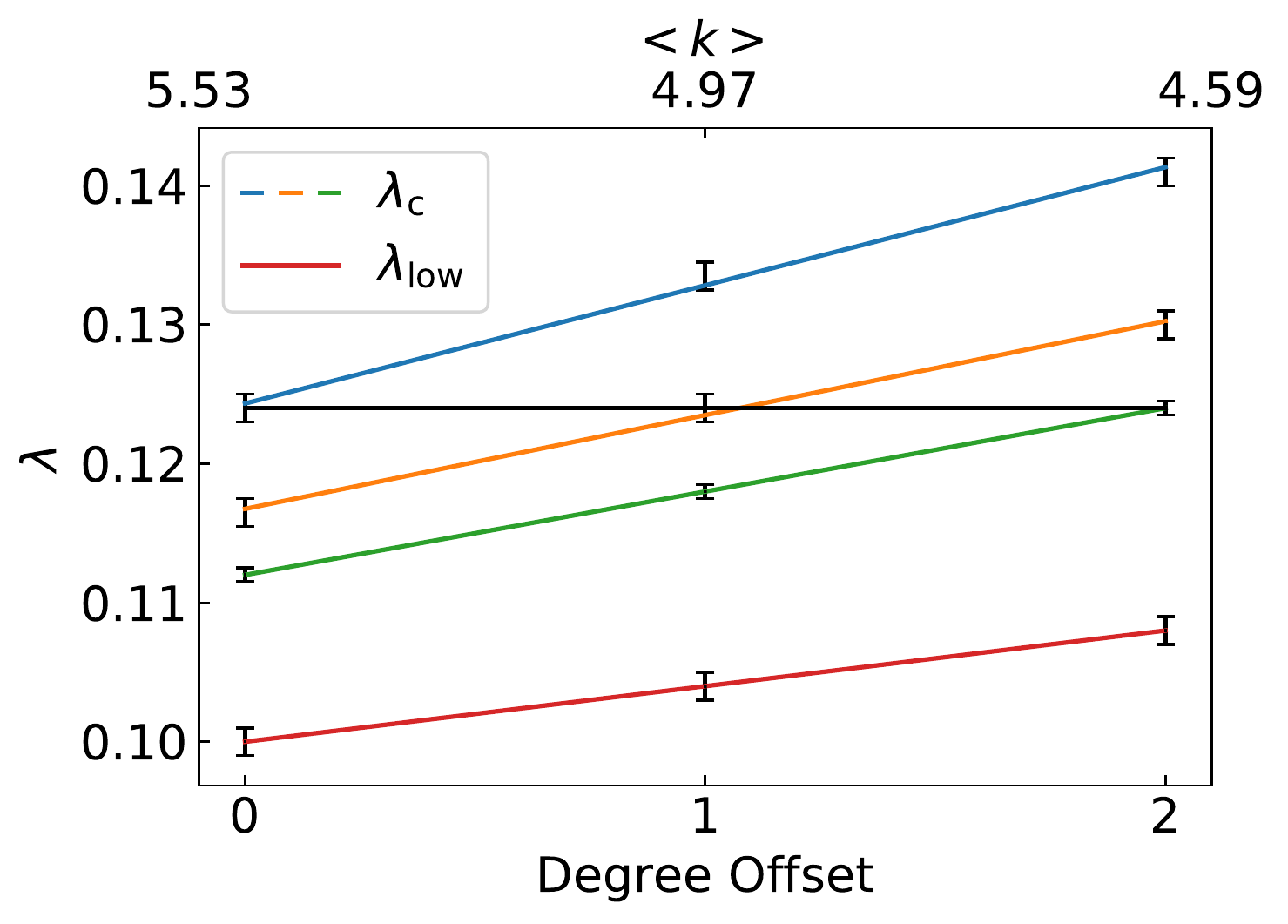}%
	}
	\subfloat[\label{fig:gp_over_gamma}]{%
		\includegraphics[width=0.5\linewidth]{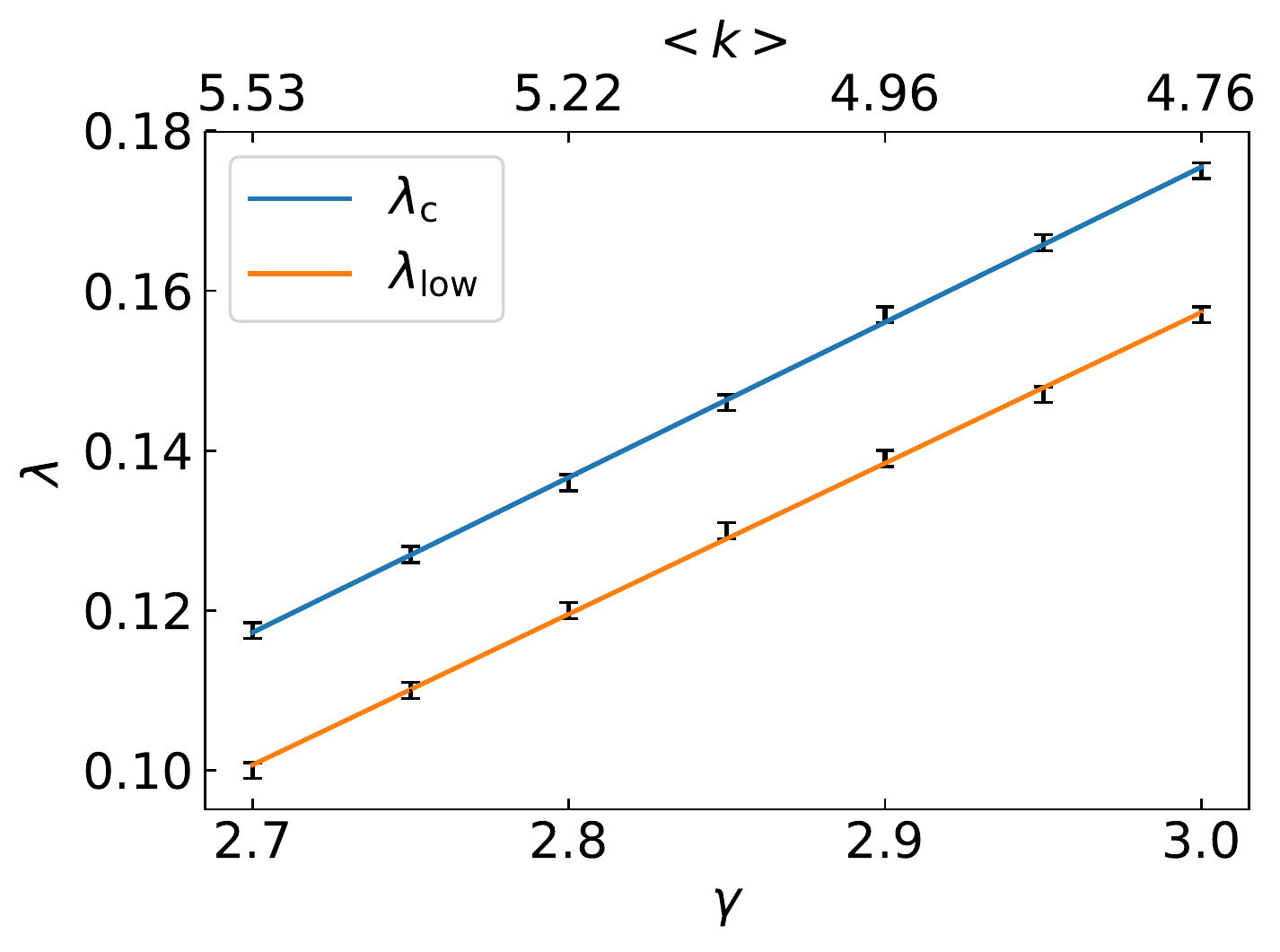}%
	}
    \caption{The upper and lower limit of $\Delta$GP in modular power-law networks with ${M=N=10^3}$ modules and nodes at (a) ${k_{\mathrm{inter}} = 3,11,38}$ (top to bottom lines) and (b) ${k_{\mathrm{inter}} = 10}$ with decreasing intramodular connectivity. (a) Reducing the degree of each node leads to an increase in $\Delta$GP as $\lambda_{\mathrm{c}}$ increases steeper than $\lambda_{\mathrm{low}}$. The black bar marks equal critical point with dynamics showcased in Figure~\ref{fig:counter_tuning}. (b) Increasing the power-law exponent $\gamma$ of the degree distribution reduces the degree of outlier nodes. It keeps $\Delta$GP constant and moves the Griffiths phase to a different parameter region. $\lambda_{\mathrm{c}}$ scales as ${\sim 1/\sqrt{k_{\mathrm{max}}}}$, in accordance with quenched mean-field theory (QMF) \cite{Castellano_2010}.}
	\label{fig:gp_over_intra}
\end{figure}

Our second approach to lower the intramodular connectivity is to increase the power-law exponent of the degree distribution $\gamma$. This leads to a lower chance to draw high-degree nodes, which reduces connectivity via the outlier nodes of the modules. This method leads to an increase in $\lambda_{\mathrm{c}}$ and $\lambda_{\mathrm{low}}$ at the same rate, which moves the Griffiths phase to a different parameter region (cf. Figure~\ref{fig:gp_over_gamma}). Both methods increase $\lambda_{\mathrm{c}}$ and reduce the average degree $\langle k \rangle$, but since a higher $\gamma$ reduces $\langle k \rangle$ via predominantly high degree nodes, it has a stronger influence on $\lambda_{\mathrm{c}}$ per removed link, as we can see by comparing Figure~\ref{fig:crit_over_off}b.

\subsection{Maintaining a stable critical region}

\begin{figure}
    \captionsetup[subfigure]{justification=centering}
    \centering
	\subfloat[$k_{\mathrm{inter}} = 3$, offset-0\label{fig:mod_pl0_c3_decay}]{%
		\includegraphics[width=0.33\linewidth]{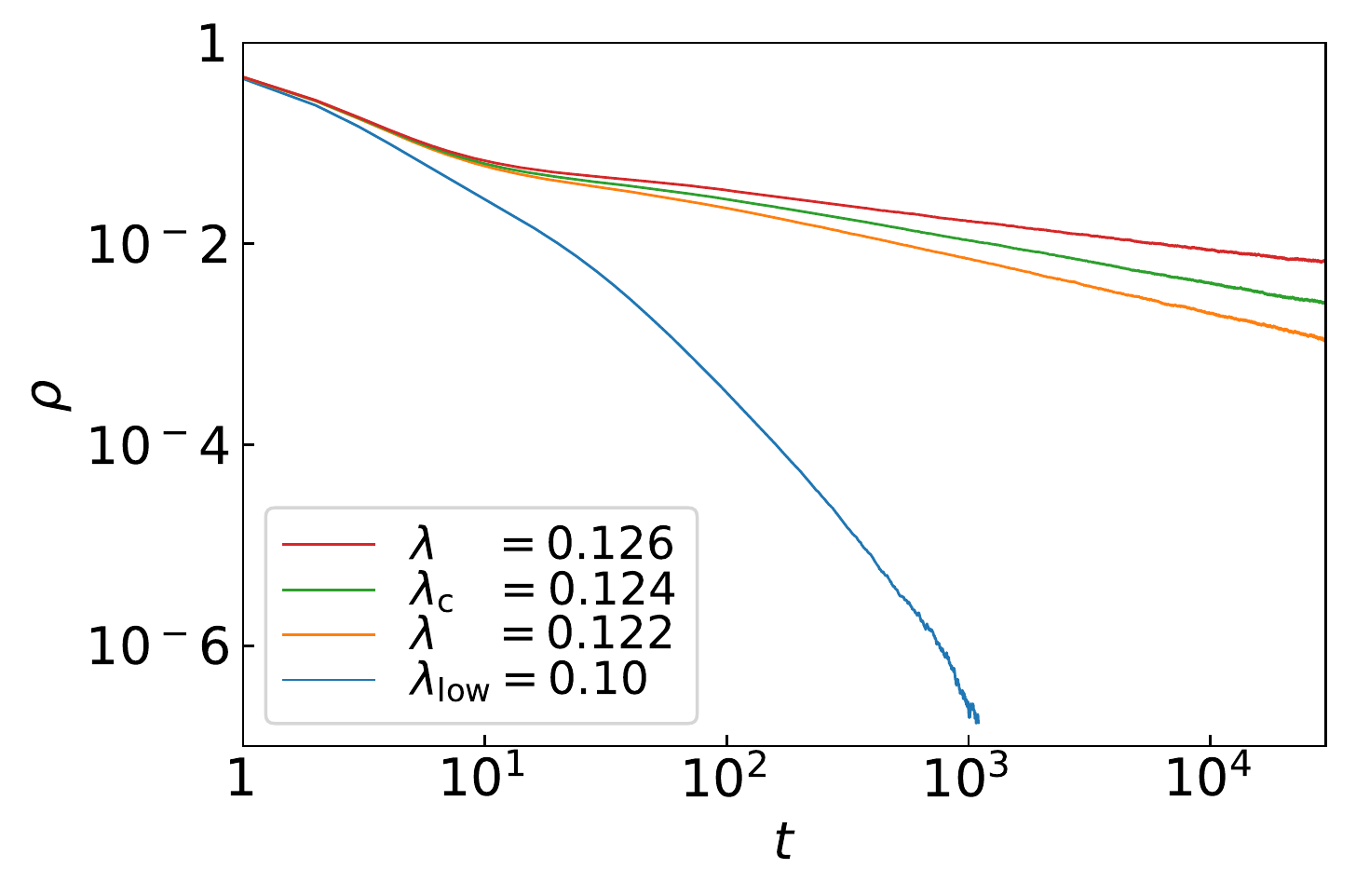}%
	}
	\subfloat[$k_{\mathrm{inter}} = 11$, offset-1\label{fig:mod_pl1_c12_decay}]{%
		\includegraphics[width=0.33\linewidth]{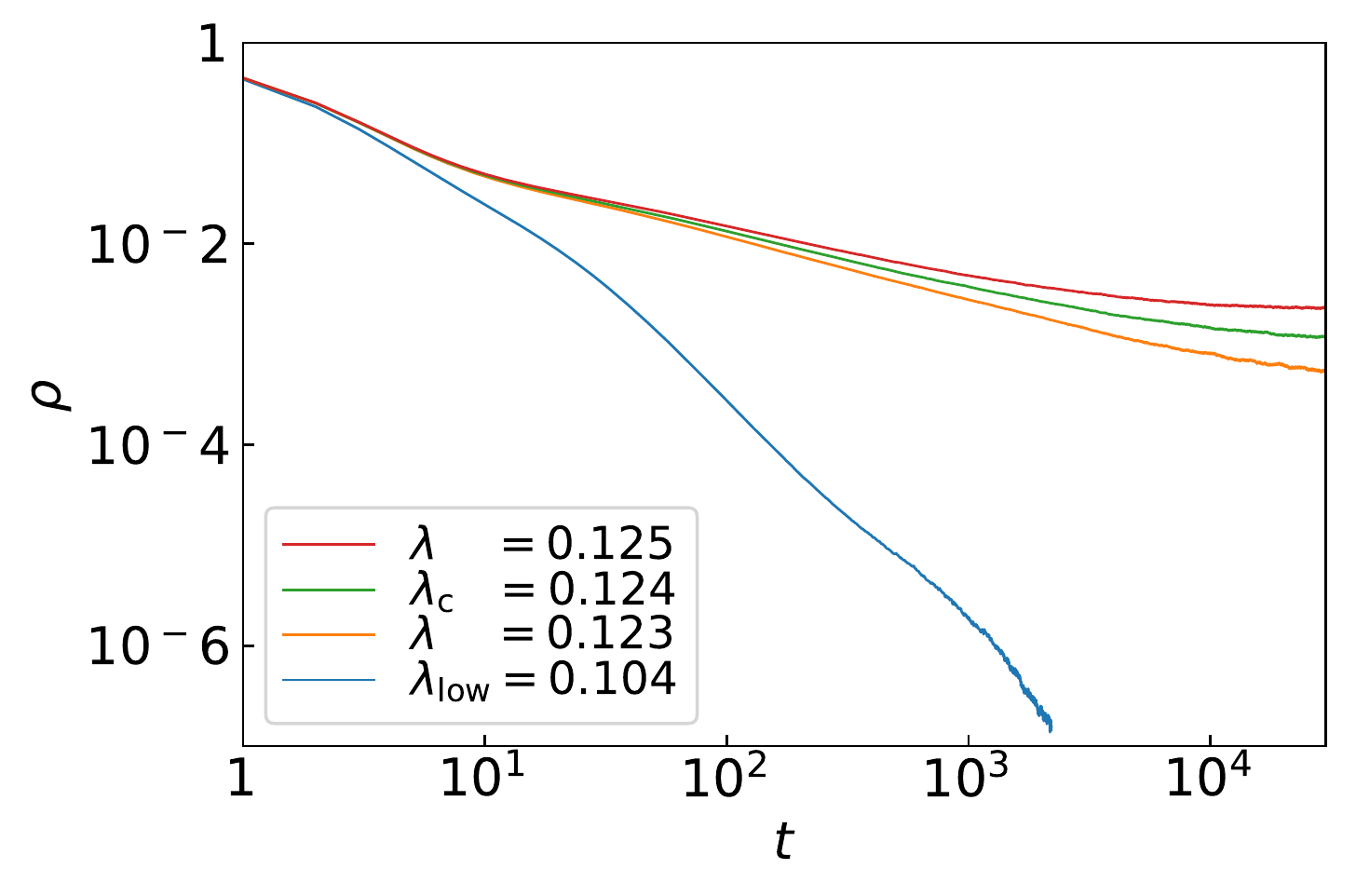}%
	}
	\subfloat[$k_{\mathrm{inter}} = 38$, offset-2\label{fig:mod_pl2_c38_decay}]{%
		\includegraphics[width=0.33\linewidth]{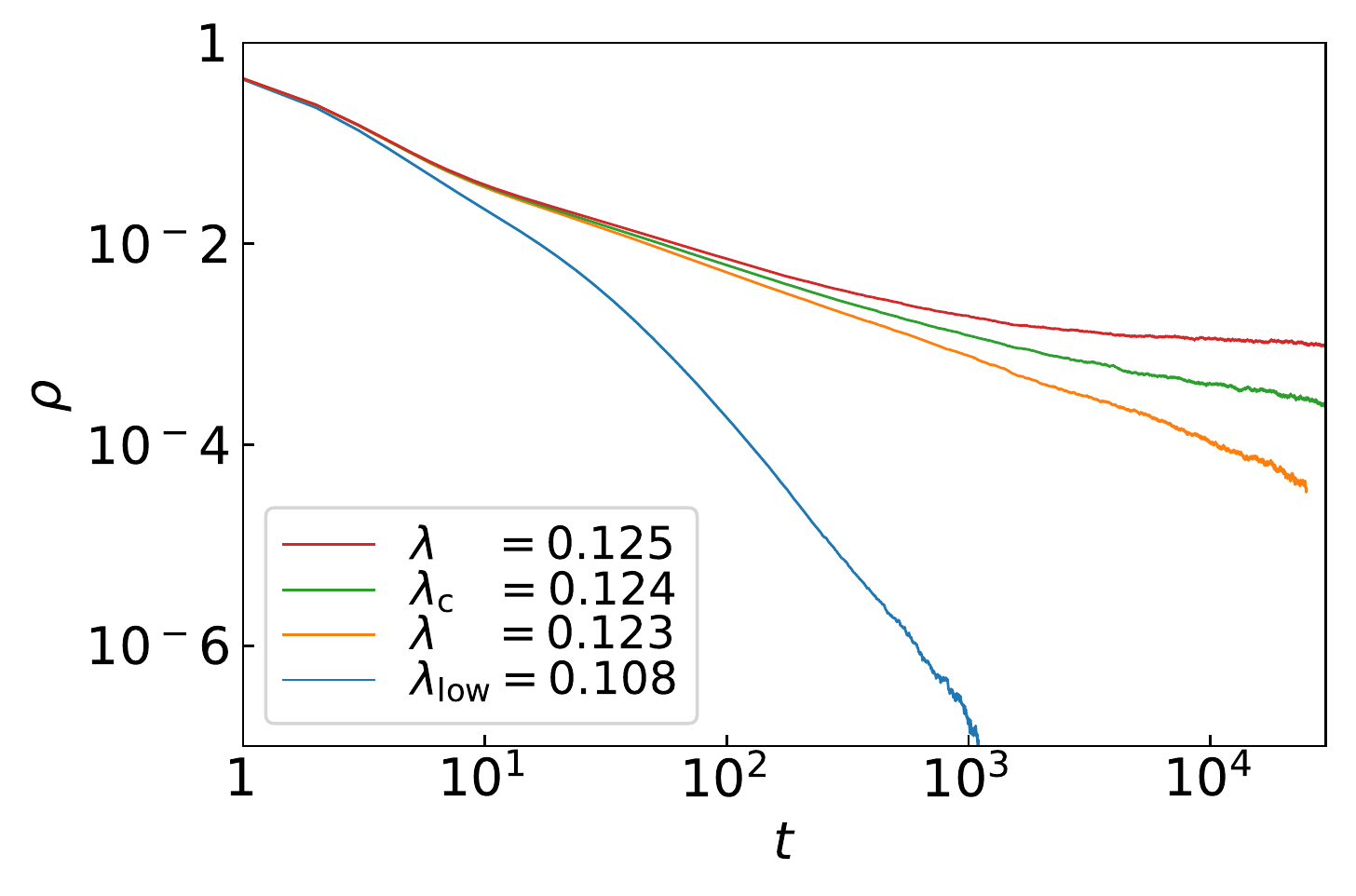}%
	}
	\caption{Activity density decay at the upper and lower limit of the Griffiths phase in modular power-law networks with exponent ${\gamma = 2.7}$ with ${M=N=10^3}$ modules and nodes and intra- and intermodular connectivity as marked by the black lines in Figure~\ref{fig:mod_decay_intra} and \ref{fig:crit_over_off}. The networks have intra- and intermodular connectivity tuned contrary to each other and have the same $\lambda_{\mathrm{c}}$ and similar power-law decay regions. A higher offset leads to an increased $\lambda_{\mathrm{low}}$ and decreased $\Delta$GP, a higher $k_{\mathrm{inter}}$ leads to faster decay to steady states above $\lambda_{\mathrm{c}}$ and shorter lifetimes below. From left to right the networks global efficiency increases and entropy decreases, as marked in Figure~\ref{fig:eff_over_k1} and \ref{fig:geoent_kinter}.}
	\label{fig:counter_tuning}
\end{figure}

Figure~\ref{fig:mod_decay_intra} shows that network configurations with different connectivity can have equal values of either $\lambda_{\mathrm{c}}$ or $\Delta$GP, if $k_{\mathrm{inter}}$ and offset are tuned accordingly. In Figure~\ref{fig:counter_tuning} we see the density decay of three networks with varying topology tuned to equal $\lambda_{\mathrm{c}}$. The networks at higher offsets have an increased $\lambda_{\mathrm{low}}$, and therefore reduced $\Delta$GP, but the three networks still have a significant overlap in power-law decay region, despite highly varying global efficiency and both geodesic and degree entropy (cf. Figure~\ref{fig:eff_over_k1},~\ref{fig:geoent}).

\subsection{Measuring the topological changes}

\begin{SCfigure}
	\centering
	\includegraphics[width=0.5\linewidth]{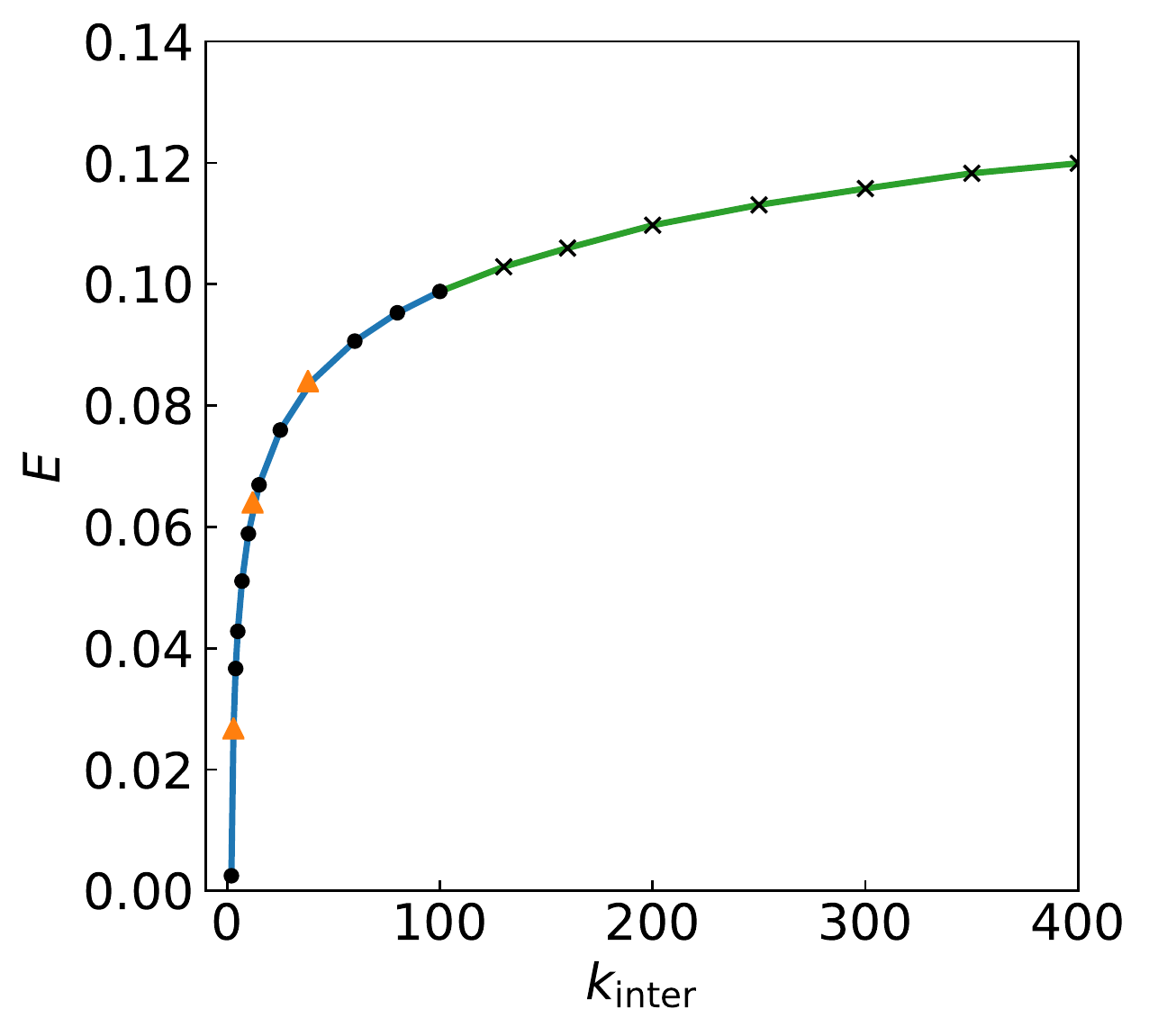}
	\llap{\shortstack{%
			\includegraphics[scale=.3]{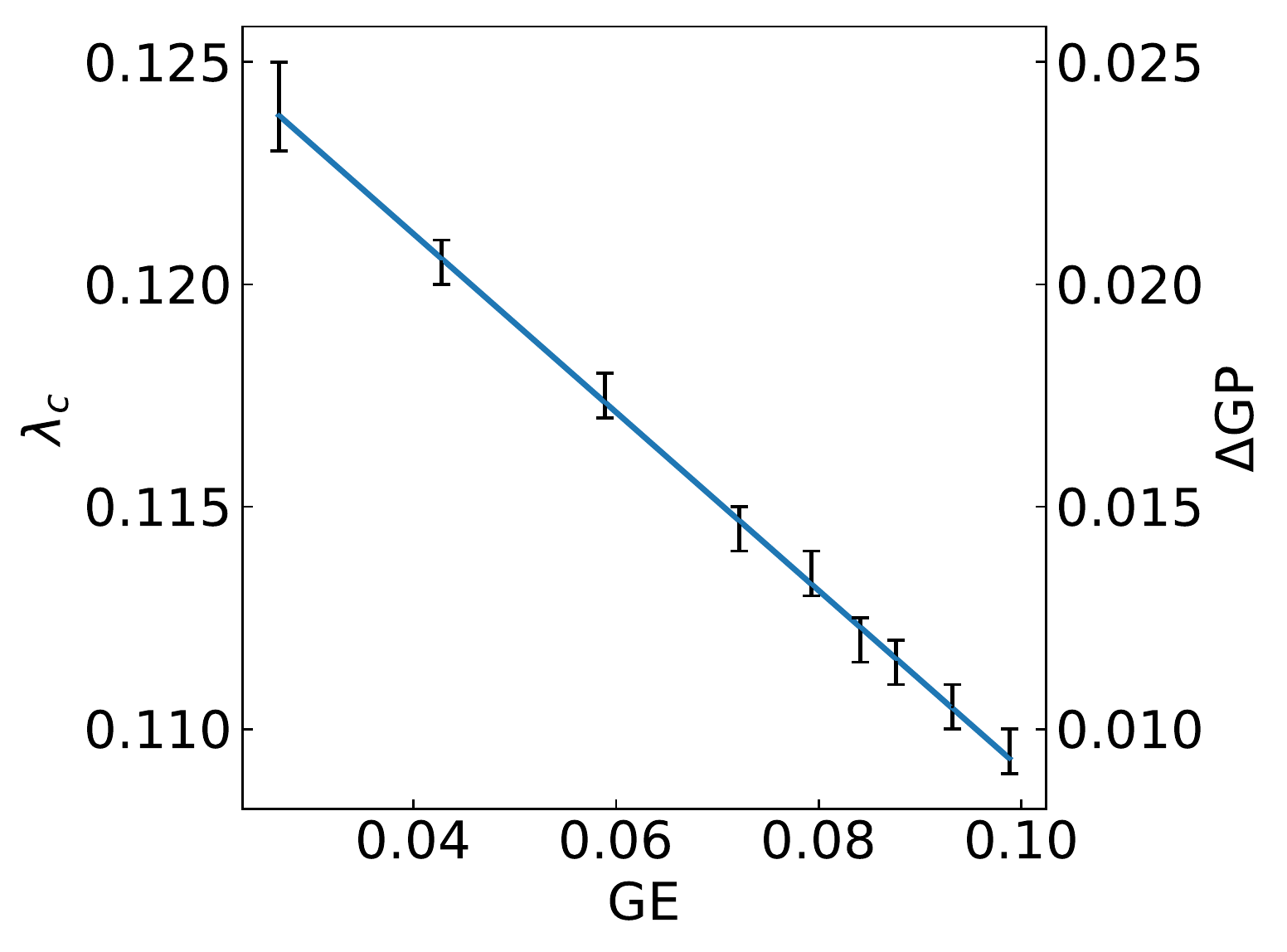}\\
			\rule{0ex}{0.4in}%
		}
		\rule{0.3in}{0ex}}
	\caption{Global efficiency in modular power-law networks with exponent ${\gamma = 2.7}$ with ${M=N=10^3}$ modules and nodes, with increasing intermodular connectivity. At ${k_{\mathrm{inter}} = 3}$ the networks shift from disconnected rings of modules to a connected small-world, which leads to a drastic reduction in average path length. Global efficiency scales inverse to average path length and increases non-analytically from zero to a finite value. This marks the presence of a percolation phase transition \cite{Dorogovtsev2008} with global efficiency as the order parameter. The coloring distinguishes the values at which extended power-law decay was observed (blue) and where it transitions into exponential decay (green). The density decay at the orange triangles in is shown in Figure~\ref{fig:counter_tuning}.}
	\label{fig:eff_over_k1}
\end{SCfigure}

Increasing intermodular connectivity leads to a decrease in the average shortest path length between nodes and an increase in global efficiency. Figure~\ref{fig:eff_over_k1} depicts the relation between global efficiency, intermodular connectivity and the Griffiths phase. The increase of global efficiency due to the increase in intermodular connectivity is particularly strong at low intermodular connectivity, when the modular networks are close to segregation. The reason for that lies in the presence of a percolation phase transition at ${k_{\mathrm{inter}}=3}$ \cite{Dorogovtsev2008}. We observe that global efficiency can serve as the order parameter of the percolation phase transition. Close to ${k_{\mathrm{inter}}=3}$ global efficiency is therefore highly sensitive to changes in $k_{\mathrm{inter}}$. The inset in Figure~\ref{fig:eff_over_k1} shows that this sensitivity extends to $\lambda_{\mathrm{c}}$ and $\Delta$GP, as they scale linearly with global efficiency.

\begin{figure}
    \centering
	\subfloat[]{\label{fig:geoent_kinter}%
		\includegraphics[width=0.50\linewidth]{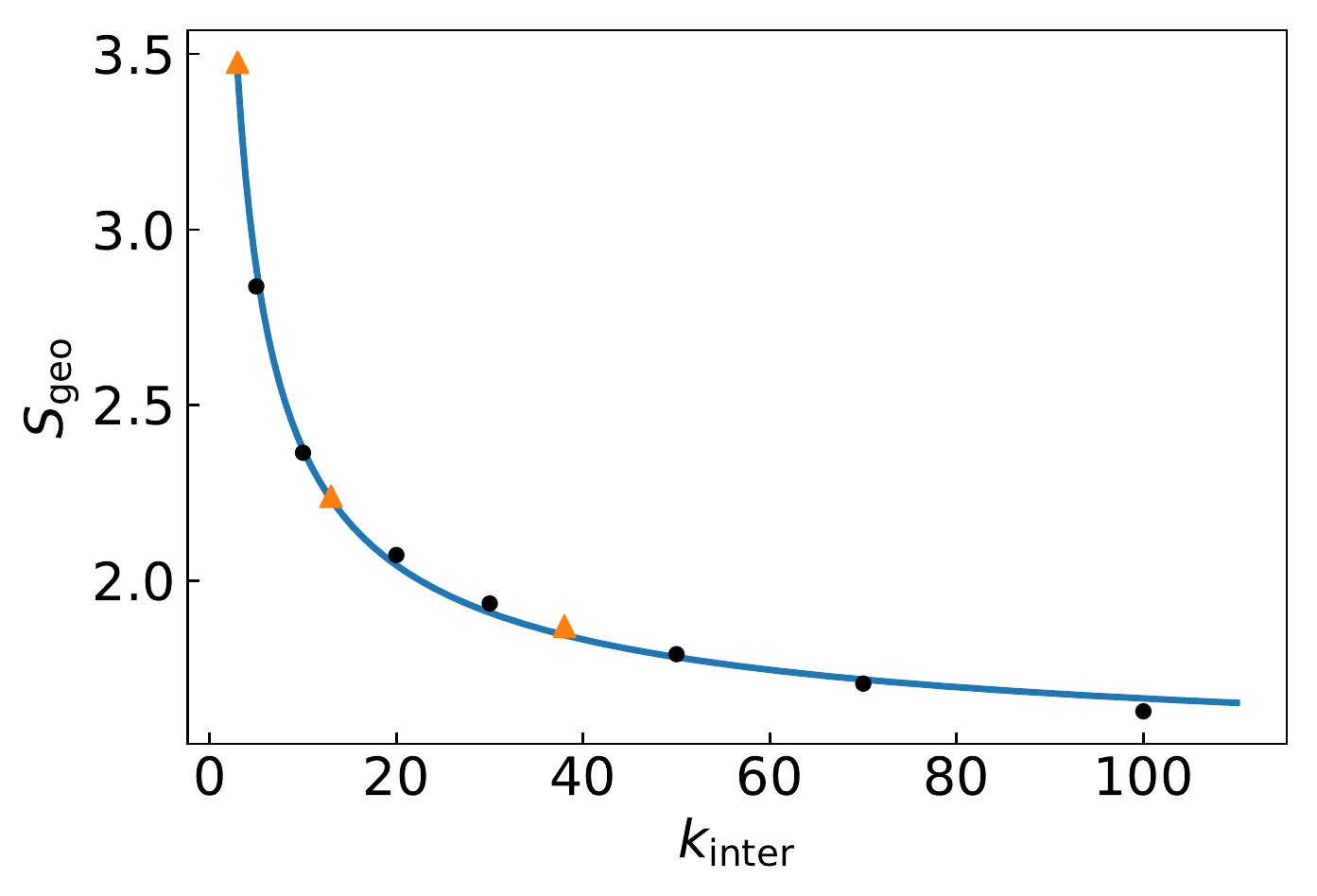}
		\llap{\shortstack{%
			\includegraphics[scale=.28]{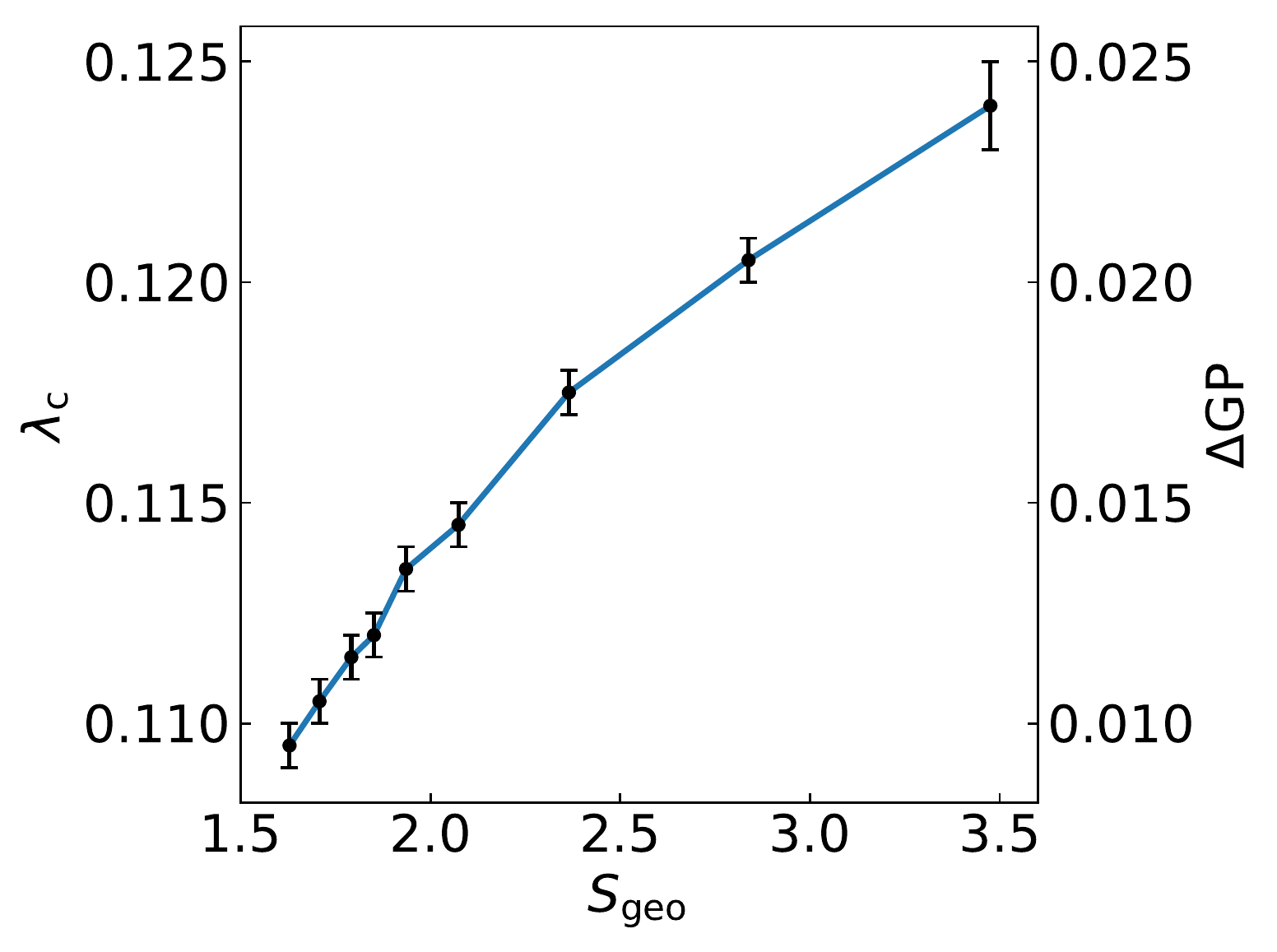}\\
			\rule{0ex}{0.67in}%
		}
		\rule{0.15in}{0ex}}%
	}
	\subfloat[]{\label{fig:geoent_intra}%
	    \includegraphics[width=0.50\linewidth]{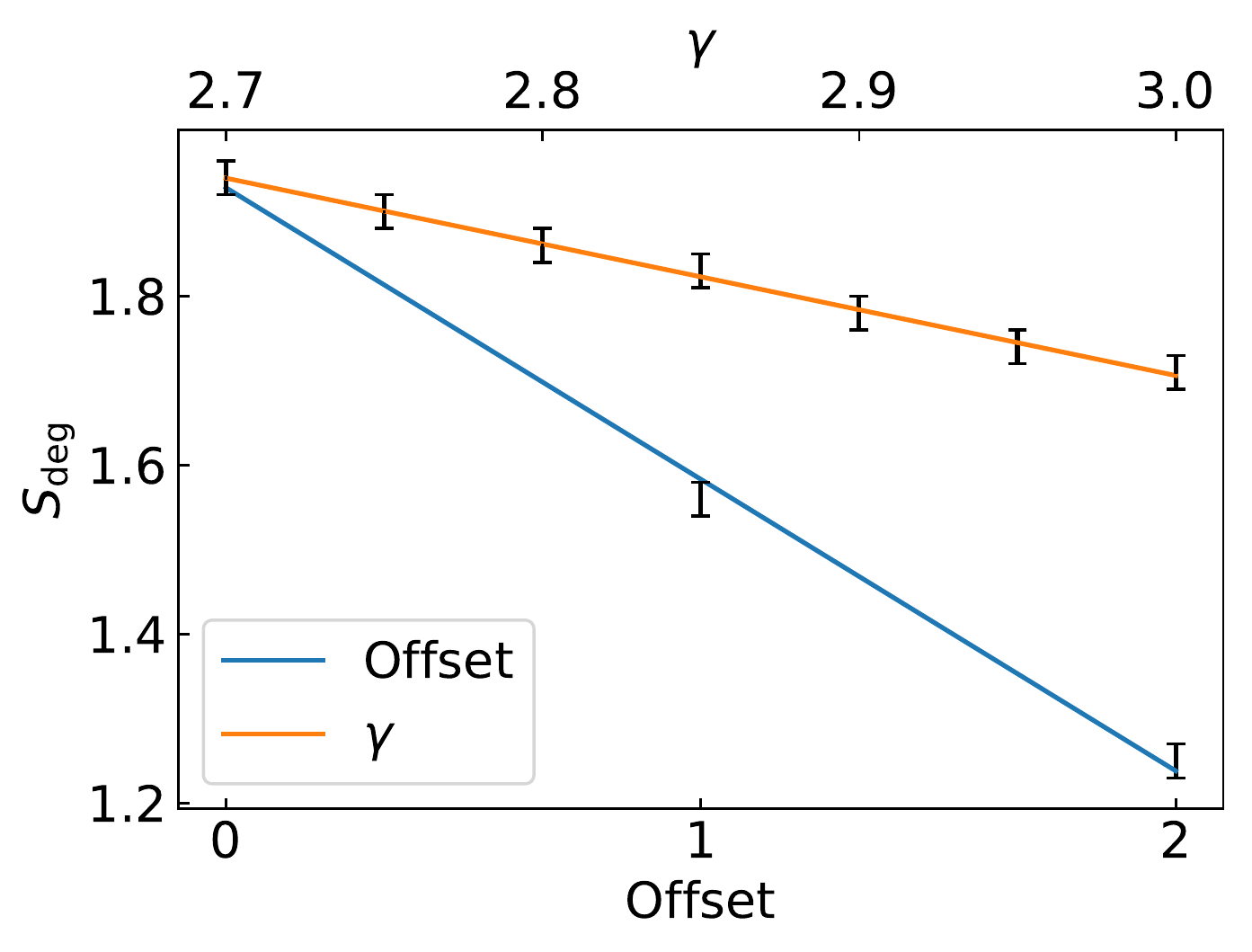}%
	}
	\caption{(a) Geodesic entropy at ${\gamma=2.7}$ with increasing intermodular connectivity and (b) degree entropy with decreasing intramodular connectivity in modular power-law networks with ${M=N=10^3}$ modules and nodes. Geodesic entropy is not significantly affected by changes in intramodular connectivity and degree entropy is independent of intermodular connectivity. The density decay at the orange triangles in (a) is shown in Figure~\ref{fig:counter_tuning}.}
	\label{fig:geoent}
\end{figure}

Geodesic entropy is also connected to the average path length and decreases with increasing intermodular connectivity, as the variability of shortest paths decreases (cf. Figure~\ref{fig:geoent_kinter}). In the inset plot we can see that in modular networks a higher geodesic entropy correlates with a larger Griffiths phase width. Degree entropy on the other hand is higher with larger intramodular connectivity (cf. Figure~\ref{fig:geoent_intra}).

\begin{SCfigure}
	\centering
	\includegraphics[width=0.5\linewidth]{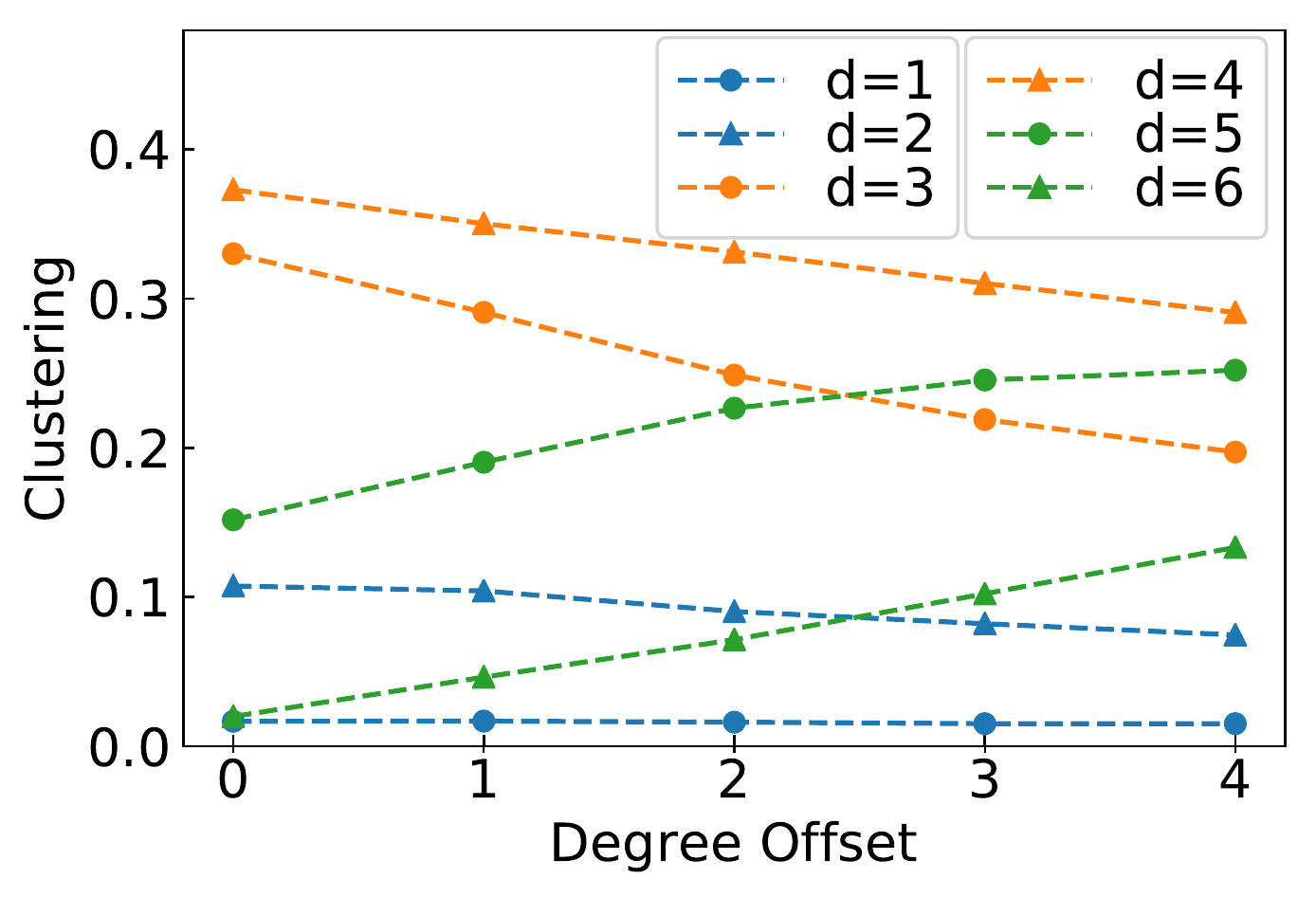}
	\caption{Extended clustering in modular power-law networks with exponent ${\gamma = 2.7}$ with ${M=N=10^3}$ modules and nodes at ${k_{\mathrm{inter}}=3}$ and decreasing intramodular connectivity. The standard averaged local clustering coefficient (${d=1}$) implies no clustering and remains unchanged with rising degree offset. Yet deeper measures (${d=3,4}$) reveal a clustered structure that shifts to a more distant clustering (${d=5,6}$) with rising degree offset.}
	\label{fig:ext_clustering}
\end{SCfigure}

Despite a reduction in average degree, when intramodular connectivity is reduced, local clustering and efficiency remain unchanged. This indicates that local clustering is not responsible for the changes observed in the Griffiths phase. We therefore evaluate the extended clustering coefficient \cite{Abdo2006}, which is a generalization of the traditional local clustering coefficient. While the local clustering coefficient measures connectivity in the direct neighborhood of a node, the extended clustering coefficient can additionally detect clusters of greater distance. Figure~\ref{fig:ext_clustering} shows that the reduction of intramodular connectivity increases the distance of extended clustering and therefore leads to a less clustered structure. A change in intermodular connectivity does not affect extended clustering. On the other hand, intramodular connectivity does not affect overall efficiency.

\section{\label{Discussion}Discussion}

\noindent
The detection of a Griffiths phase in complex networks appears to be an important step towards understanding critical behavior in biological systems \cite{Moretti2013}. An extended critical region in brain-like networks supports the hypothesis that the brain operates at criticality \cite{Beggs2008} and relaxes the necessity for fine-tuning parameters around a precise critical point. Previous work demonstrates that structural heterogeneity and modularity, both features of human brain networks \cite{Meunier2010, Sporns2010, Lopez2016, Gallos2012}, are sufficient conditions to enable a Griffiths phase \cite{Cota2018}. Functional brain networks have been shown to change significantly in different mind conditions, such as sleep, coma, anesthesia or under the influence of psychedelics \cite{Viol2019,Noirhomme2010, DeAraujo2012, Carhart-Harris2012, Palhano-Fontes2015}. Studies indicate a neural correlate between brain network integration and consciousness states \cite{Andrade2011, Schrouff2011, Carhart-Harris2016, Luppi2019, Luppi2021} which some quantify using topological network metrics such as the clustering coefficient, local and global efficiency and entropy \cite{Viol2017, Schroter2012, Viol2019}.  We argue whether the observed changes in topological metrics, especially during psychedelic states, can influence critical behavior by altering the expression of an existing Griffiths phase.

We construct modular networks with varying intra- and intermodular connectivity and explore numerically how the connectivity structure affects the critical behavior of a spreading process and evaluate the link between the observed changes and topological network metrics. We show that an increase in either connectivity leads to a decline of the networks critical point, hence a reduction of the Griffiths phase width. In addition, a decrease in intramodular connectivity leads to a slight increase in the lower limit of the Griffiths phase. Intra- and intermodular connectivity therefore offer two independent ways of controlling critical dynamics and can be used as a tuning mechanism for criticality. If one connectivity structure is changed, we can adapt the opposing structure, leading to networks with differing topology, but identical critical points and similar Griffiths phase regions.

We observe that global efficiency is a central measure in the emergence of a Griffiths phase. The Griffiths phase width scales linearly with global efficiency in the modular networks. Global efficiency captures the reduced information exchange at low intermodular connectivity which enables rare regions effects.

We find that a decreased intermodular connectivity leads to an increase in entropy, even more so when intramodular connectivity is adapted to counter changes in criticality. Recent researches suggests that psychedelics disrupt the hierarchy of brain network topology \cite{barnett2020,alonso2015} and increase network segregation, indicated by increased shortest path length \cite{Viol2017,Luppi2021} and decreased global efficiency \cite{Viol2017}. We argue whether these networks would exhibit a larger Griffiths phase width, similar to the synthetic networks we present here, and whether this increase could explain the enhanced entropy observed in functional brain networks of individuals under psychedelic influence \cite{Viol2017,Viol2019,Carhart-Harris2016,Tagl2014}.

We hypothesize that the topological changes observed in functional human brain networks during altered states of consciousness may be connected to a mechanism that ensures critical operation. Either by tuning topological parameters to maintain a stable critical region or by extending the critical region. An increased Griffiths phase width via a heightened critical point moves activity rates that were previously supercritical to the critical region. This increases the dynamical parameter range in the networks and could explain the increase of entropy observed during psychedelic states. Further studies should clarify our hypothesis by reproducing these results in networks with more empirical parameters.

We highlight some limitations of our work. The networks are dynamically heterogeneous with the SIS due to the power-law distribution within the modules. Different propagation models, such as the contact process, would require a different intramodular structure to achieve a distribution of critical points among the modules \cite{Cota2018} and the structure manipulation would have to be adapted. We expect, however, that the modulation of the critical point is reproducible in any network with distinct intra- and intermodular structures and a spreading process that can be separately affected by intramodular properties and changes in average path length created through alterations in intermodular connectivity. Intramodular connectivity can be altered via either low or high degree nodes. Decreasing the degree of each by node by a chosen value affects predominantly low degree nodes and leads to an increase in the Griffiths phase width. This method is only appropriate for small off-sets because large values can distorts the power-law distribution. By increasing the power-law exponent of the degree distribution we focus the connectivity reduction to high degree nodes. This keeps the Griffiths phase width constant and moves the Griffiths phase to a different parameter region. Further research could be focused on how both width and parameter region of the Griffiths phase could be continuously controlled via the intramodular distribution.

An interesting point of further study is if our results are applicable to the spread of infectious diseases, such as the current outbreak of SARS-CoV-2. 
If we can model the social networks of international communities as sufficiently heterogeneous modular networks, there is a possibility for the emergence of Griffiths effects. This could explain unexpectedly long lifetimes of spreading processes, despite a sub critical spreading rate.

\section{Conclusion}

To conclude, we show that the extension of a critical region in modular networks can be controlled via both intra- and intermodular connectivity individually. Changes in critical dynamics that stem from a change in either connectivity can be counteracted by tuning the opposing structure. We find that low global efficiency is central in the emergence of a Griffiths phase and connect our results to the observed changes in functional brain networks in altered states of consciousness.

\newpage

\appendix

\section{Computational details}

\subsection{Generating the modular networks\label{appendix_networks}}

The networks considered in this paper are generated through the following steps:
\begin{enumerate}[label=(\roman*)]
	\item $M$ connected modules containing $N$ nodes are generated via the uncorrelated configuration model, drawing from a degree distribution $P(k)$. 
	\item Multi- and self-edges are randomly rewired within the modules.
	\item $k_{\mathrm{inter}}$ stubs are selected in each module during the generation process and preserved for linking to other modules, so that the overall degree distribution remains unchanged.
	\item The preserved stubs are randomly linked to each other. Linking of stubs within the same module is prohibited.
	\item The process is iterated until the whole network is fully connected.
\end{enumerate}

\subsection{SIS implementation\label{appendix_SIS}}

During the simulation we keep track of the infected/active nodes, their total number $N_{\mathrm{inf}}$ and their degrees $k_{\mathrm{inf}}$.
\begin{enumerate}[label=(\roman*)]
	\item  With probability
	\begin{equation}
	    p = \frac{\mu N_{\mathrm{inf}}}{\mu N_{\mathrm{inf}} + \lambda k_{\mathrm{inf}}}
	\end{equation}
	a randomly selected active node becomes inactive. For simplicity, we fix the recovery rate at ${\mu = 1}$.
    \item With probability $1-p$ an active node is selected with a probability proportional to its degree ${k/k_{\mathrm{max}}}$. A randomly chosen neighbor of the node is selected and, if inactive, becomes active. If it is already active, the simulation continues with (iii).
    \item Time is incremented by
    \begin{equation}
	    \tau = \frac{\mathrm{ln}(u)}{\mu N_{\mathrm{inf}} + \lambda k_{\mathrm{inf}}},
	\end{equation}
	where $u$ is a pseudo random number, uniformly distributed in the interval ${(0, 1)}$. The steps are repeated $t$ times or until ${N_{\mathrm{inf}} = 0}$.
\end{enumerate}

\newpage

\section*{Supplementary Material}

\begin{figure}[h]
    \centering
	\includegraphics[width=0.7\linewidth]{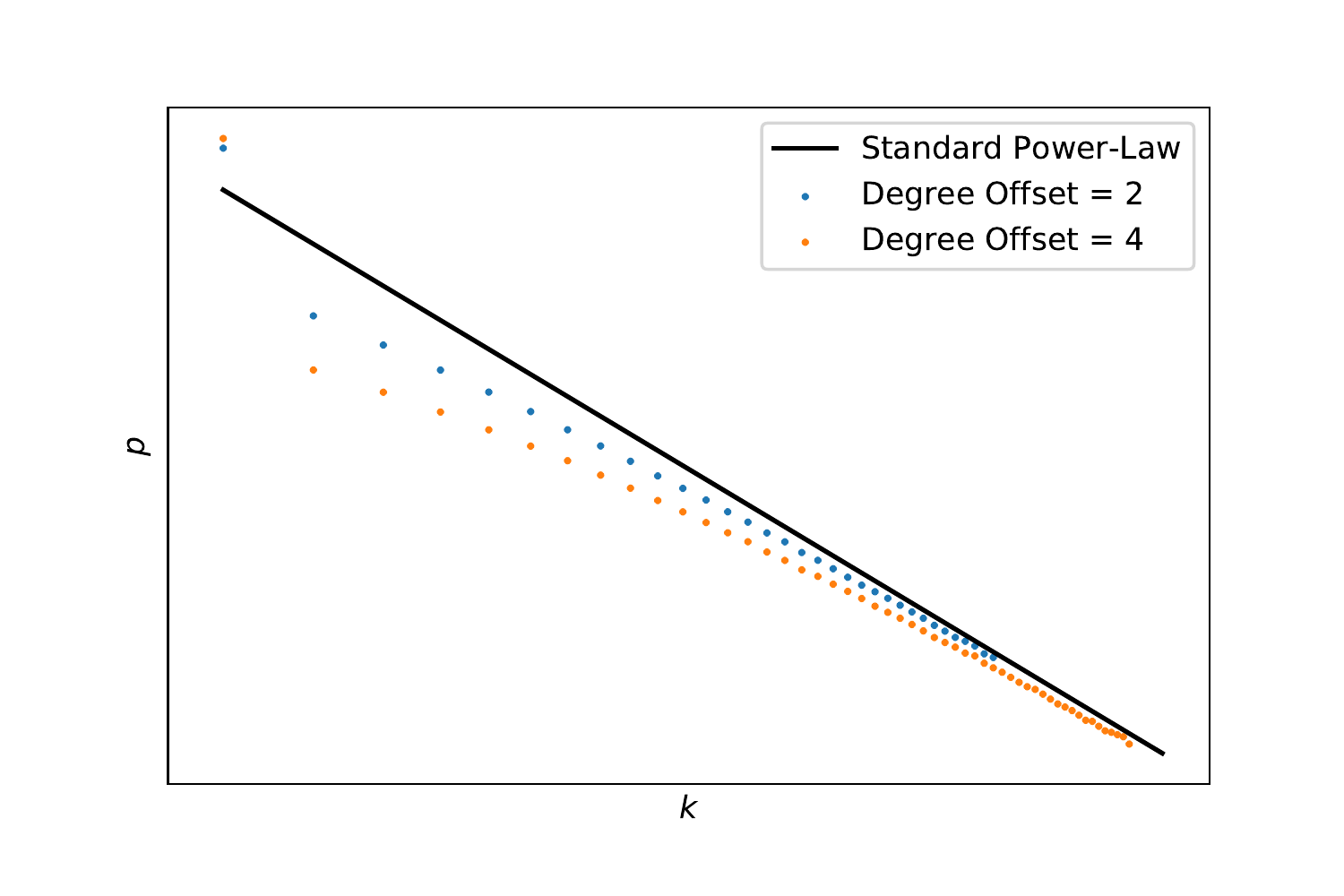}
	\caption{Degree distribution of power-law modules with exponent ${\gamma = 2.7}$ of size ${N=10^3}$ at different offset values. A degree offset in a standard power-law distribution creates an asymptotical power-law with an increased number of degree 3 nodes that asymptotically converges towards a standard power-law distribution.}
	\label{fig:asym_power_law}
\end{figure}

\clearpage

\begin{figure}
    \centering
	\subfloat[\label{fig:sup_decay_a}]{%
		\includegraphics[width=0.5\linewidth]{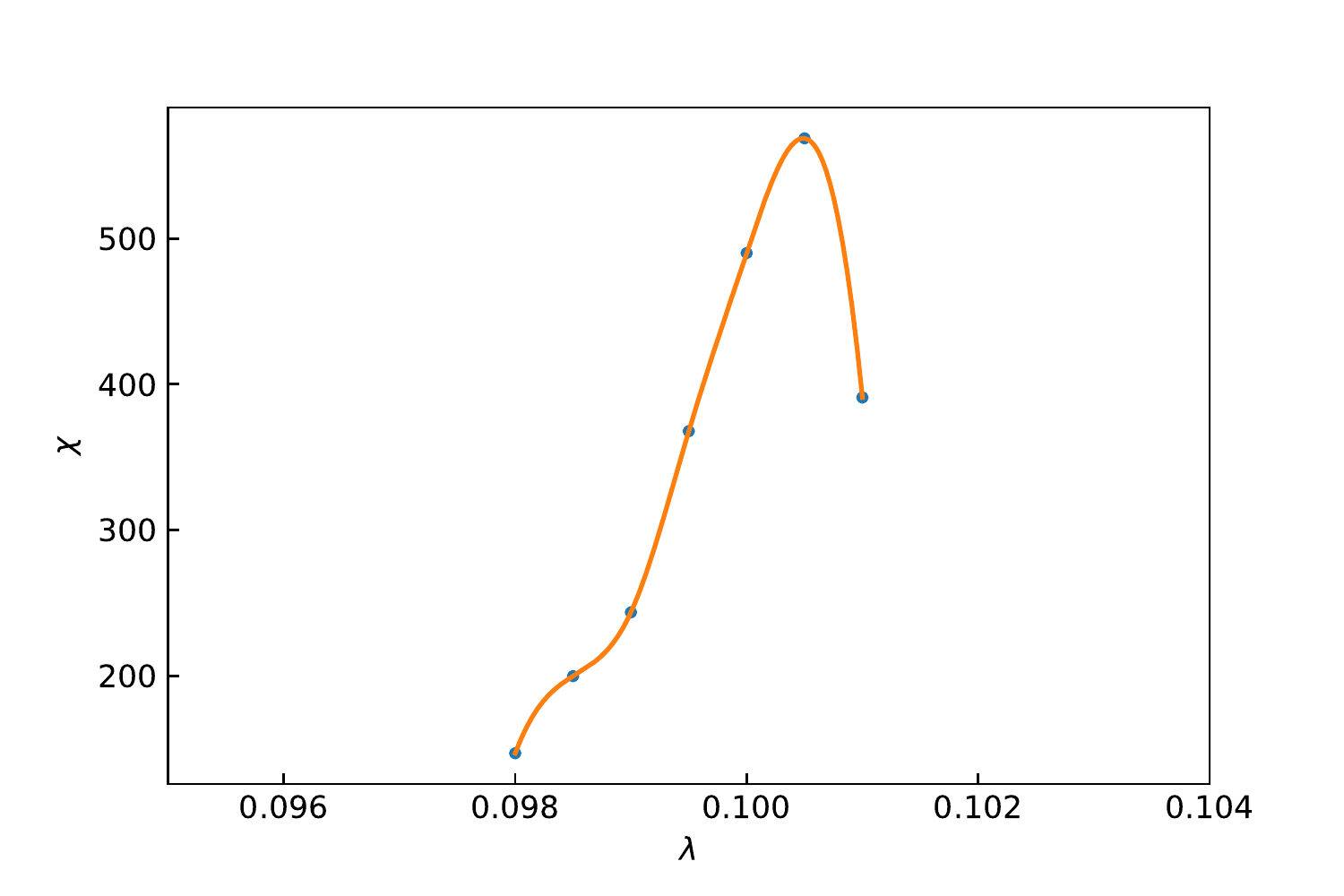}%
	}\\
	\subfloat[\label{fig:sup_decay_b}]{%
		\includegraphics[width=0.5\linewidth]{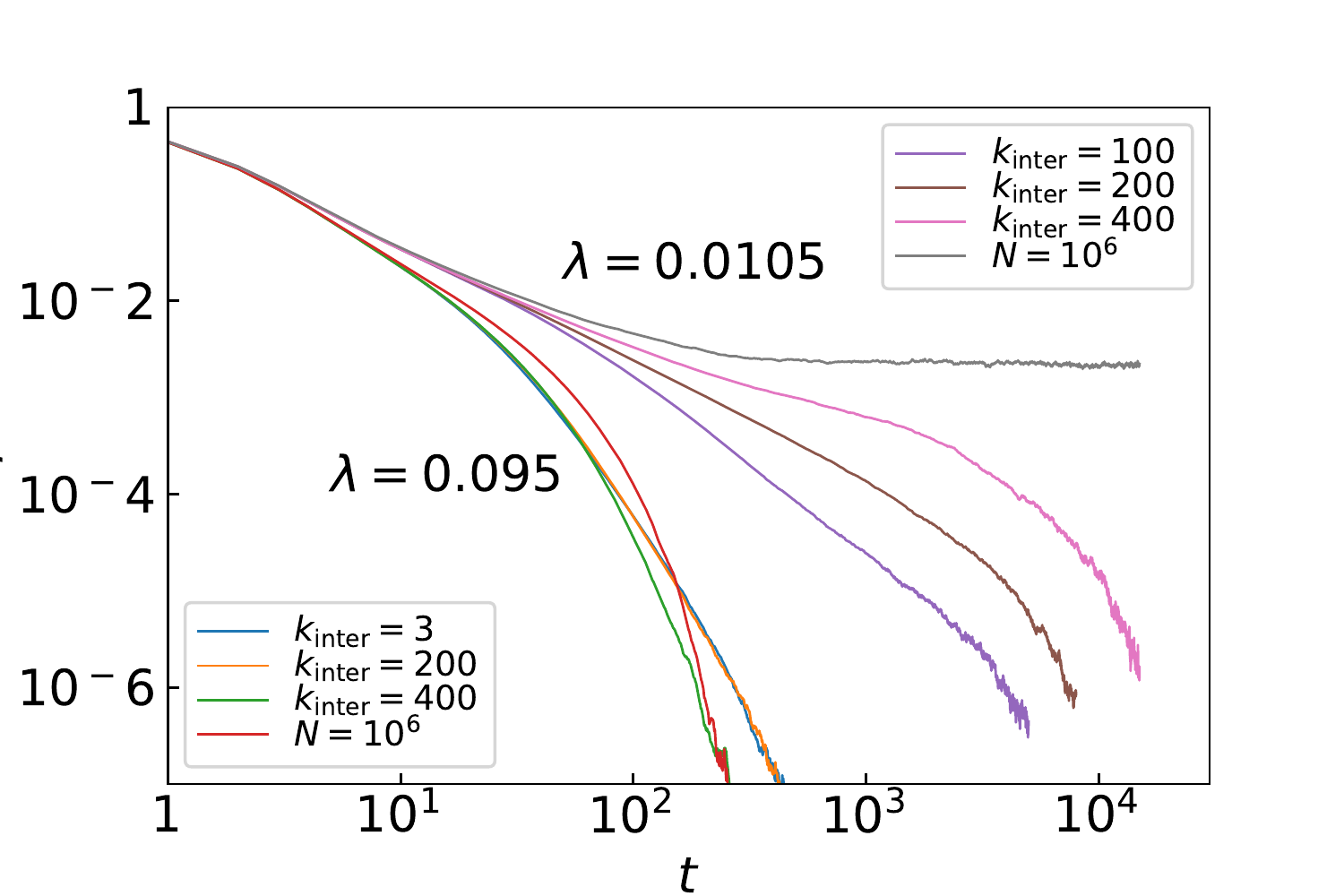}%
	}
	\subfloat[\label{fig:sup_decay_c}]{%
		\includegraphics[width=0.5\linewidth]{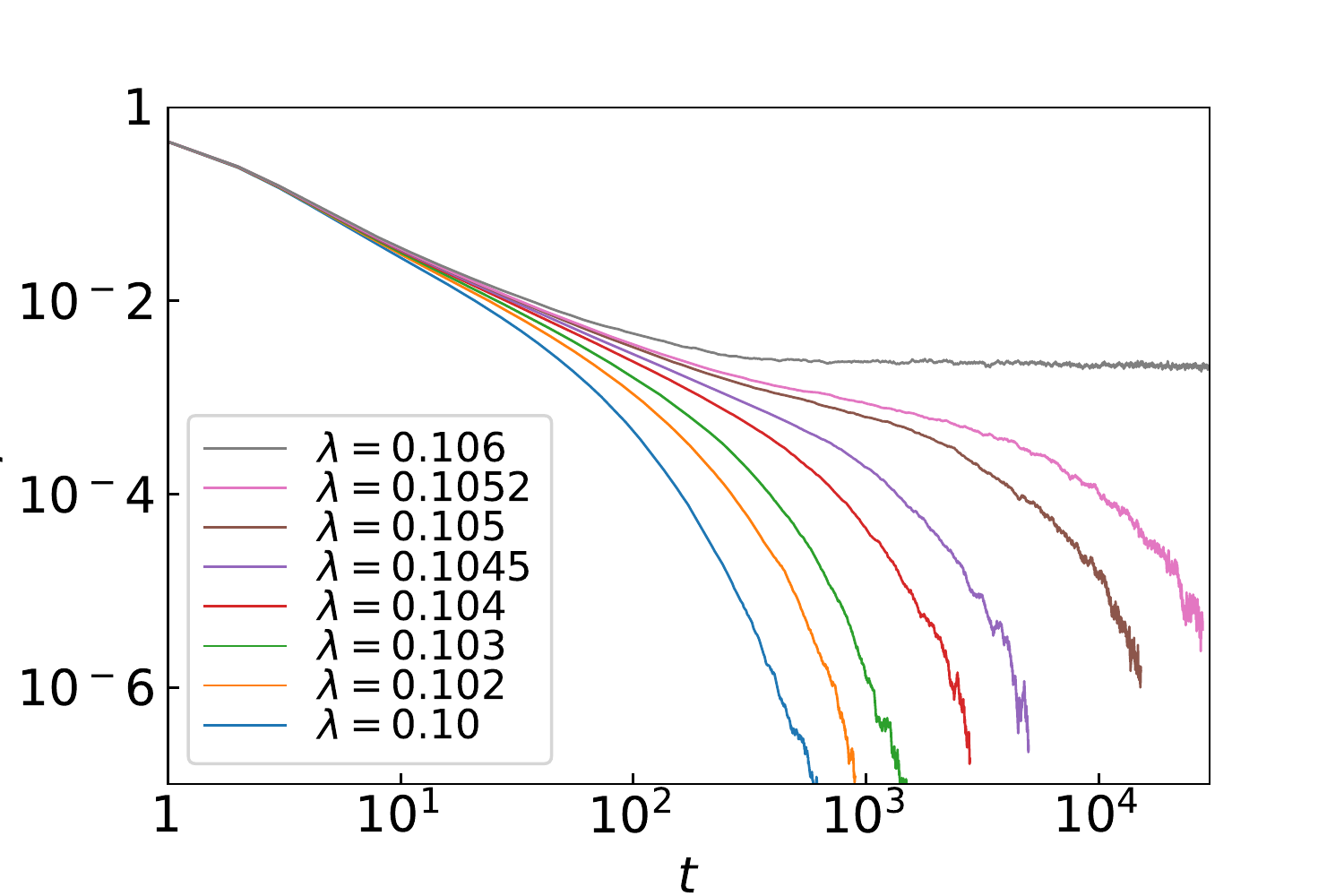}%
	}
	\caption{(a) The peak in susceptibility corresponds to the critical point ${\lambda^{'}_{\mathrm{c}} \approx 0.10}$ of a single power-law module with ${\gamma = 2.7}$ of size ${N=10^6}$. (b) We compare how the density decay of modular networks with ${M=10^3}$ modules develops below and above $\lambda_{\mathrm{low}}$ with increasing $k_{\mathrm{inter}}$. Below the lower bound of the Griffiths phase ${\lambda^{'}_{\mathrm{c}} = \lambda_{\mathrm{low}} \approx 0.10}$ the lifetime of activity is not significantly affected by the increase in intermodular connectivity. Above $\lambda_{\mathrm{low}}$ the decay gradually shifts from power-law to slow-decay with an exponential cut-off and finally to a super critical steady state when the network resembles a single module. (c) At ${k_{\mathrm{inter}} = 400}$ we do not see extended power-law decay anymore.}
	\label{fig:sup_decay}
\end{figure}

\clearpage

\section*{References}

\providecommand{\newblock}{}

\end{document}